\documentclass{article}

\usepackage[english]{babel}
\usepackage{csquotes}
\usepackage{authblk}
\usepackage{booktabs}       
\usepackage[letterpaper,top=2cm,bottom=2cm,left=3cm,right=3cm,marginparwidth=1.75cm]{geometry}
\usepackage{siunitx}
\usepackage{caption}
\usepackage{subcaption}
\usepackage{amsmath, amssymb}
\usepackage{graphicx}
\usepackage{amsfonts}
\usepackage{bm}
\usepackage[colorlinks=true, allcolors=blue]{hyperref}
\usepackage[backend=biber,style=nature,sorting=none,natbib=true,maxnames=15]{biblatex} 
\usepackage{tikz}

\usepackage{enumitem}
\setlist[itemize]{nosep}

\title{Spatial structure facilitates evolutionary rescue by drug resistance}
\date{}
\author{Cecilia Fruet\textsuperscript{1,2}, Ella Linxia Müller\textsuperscript{1,2}, Claude Loverdo\textsuperscript{3}, Anne-Florence Bitbol\textsuperscript{1,2,*}}
\affil{
\textbf{1} Institute of Bioengineering, School of Life Sciences, École Polytechnique Fédérale  de Lausanne (EPFL), Lausanne, Switzerland\\
\textbf{2} SIB Swiss Institute of Bioinformatics, Lausanne, Switzerland\\
\textbf{3} Sorbonne Université, CNRS, Institut de
Biologie Paris-Seine (IBPS), Laboratoire Jean Perrin (LJP), Paris, France\\
*anne-florence.bitbol@epfl.ch
}
\begin{document}

\maketitle

\begin{abstract}
Bacterial populations often have complex spatial structures, which can impact their evolution. Here, we study how spatial structure affects the evolution of antibiotic resistance in a bacterial population. We consider a minimal model of spatially structured populations where all demes (i.e., subpopulations) are identical and connected to each other by identical migration rates. We show that spatial structure can facilitate the survival of a bacterial population to antibiotic treatment, starting from a sensitive inoculum. Specifically, the bacterial population can be rescued if antibiotic resistant mutants appear and are present when drug is added, and spatial structure can impact the fate of these mutants and the probability that they are present. Indeed, the probability of fixation of neutral or deleterious mutations providing drug resistance is increased in smaller populations. This promotes local fixation of resistant mutants in the structured population, which facilitates evolutionary rescue by drug resistance in the rare mutation regime. Once the population is rescued by resistance, migrations allow resistant mutants to spread in all demes. Our main result that spatial structure facilitates evolutionary rescue by antibiotic resistance extends to more complex spatial structures, and to the case where there are resistant mutants in the inoculum. 
\end{abstract}

\section*{Author Summary}
Antibiotic resistance is a major challenge, since bacteria tend to adapt to the drugs they are subjected to. Understanding what conditions facilitate or hinder the appearance and spread of resistance in a bacterial population is thus of strong interest. Most natural microbial populations have complex spatial structures. This includes host-associated microbiota, such as the gut microbiota. Here, we show that spatial structure can facilitate the survival of a bacterial population to antibiotic treatment, by promoting the presence of resistant bacteria. Indeed, neutral or deleterious mutants giving resistance can take over small populations more easily than large ones, thanks to the increased importance of fluctuations in small populations. Resistant mutants can then spread to the whole structured population. Thus, population spatial structure can be a source of antibiotic treatment failure. This effect of spatial structure is generic and does not require environment heterogeneity.

\section*{Introduction}
Antibiotic resistance is a crucial challenge in public health \cite{who_antibiotics2, eu_ab}. Resistant bacteria emerge and spread through Darwinian evolution, driven by random mutations, genetic drift and natural selection. Mutations allow the emergence of strains adapted to challenging environments, including various antibiotic types \cite{hauert_evolutionary_2008}. These strains are selected when antibiotics are present in the environment. Critically, the evolution of antibiotic resistance can occur quickly, within days in specific conditions \cite{baym_spatiotemporal_2016}, while the development of a new antibiotic typically takes around ten years \cite{wong_estimation_2019}. In this context, it is crucial to understand what conditions favor or hinder the development and spread of antibiotic resistance.

Bacterial populations often have complex spatial structures. For pathogenic bacteria, each patient constitutes a different environment, connected through transmission events. Within a patient, different organs and tissues represent spatial environments between which bacteria can migrate \cite{she_defining_2024}. At a smaller scale, bacteria in the gut microbiota, which can be a reservoir of drug resistance, live in different spatial environments, namely the digesta, the inter-fold regions and the mucus, and some of them form biofilms \cite{donaldson_gut_2016, mccallum_gut_2023, chikina_attheright_2021}, which are dense structures of cells enclosed in a polymer-based matrix \cite{costerton_microbial_1995, watnick__biofilm_2000}. Thus, it is important to understand how the spatial structure of microbial populations impacts the evolution and spread of resistance. This question has been explored in epidemiological models in the case of viruses~\cite{Debarre07,Althouse13} and of bacteria~\cite{krieger_population_2020}, and was recently addressed in controlled experiments, for specific antibiotic resistant bacteria~\cite{verdon_spatial_2024}, but is still relatively under-explored. Spatial structure can lead to small effective population sizes, and small population sizes were recently experimentally shown to substantially impact the evolution and spread of resistance~\cite{Coates18,alexander_stochastic_2020}. More generally, the impact of spatial structure on the fate of mutants has been studied theoretically~\cite{Wright31,Kimura64,maruyama70,maruyama74,Nagylaki80,slatkin_fixation_1981,Barton93,Whitlock97,Nordborg02,whitlock2003,Sjodin05, lieberman_evolutionary_2005, Houchmandzadeh11, hauert_fixation_2014,allen_evolutionary_2017, marrec_toward_2021,yagoobi_fixation_2021, tkadlec_evolutionary_2023, abbara_frequent_2023} and experimentally~\cite{Kryazhimskiy12,Nahum15,france_relationship_2019,chakraborty_experimental_2023,kreger_role_2023}, but generally in constant environments. Furthermore, theoretical investigations have mainly focused on the fate of one mutant lineage, specifically on mutant fixation probability and fixation time.

Here, we ask how spatial structure impacts the overall fate of a bacterial population suddenly subjected to antibiotic treatment. Is the population eradicated by the drug, or does it survive by developing resistance? We address this question in a minimal model of population structure, composed of identical demes (i.e.\ subpopulations) connected to each other by identical migrations. This simple structure is known as the island model~\cite{Wright31,Kimura64}, or the clique or fully-connected graph~\cite{lieberman_evolutionary_2005,marrec_toward_2021,abbara_frequent_2023}. We assume that the population initially only comprises drug-sensitive bacteria.  
Indeed, the initial diversity of a population of pathogenic bacteria in a host is often low, as many infections begin with a small infectious dose \cite{gutierrez_virus_2012, lim_independent_2014, lythgoe_sarscov2_2021, woodward_gastric_2022, hoces_fitness_2023}. 
For simplicity, we assume that the environment is homogeneous, but note that environmental heterogeneities have a substantial effect on antibiotic resistance ~\cite{zhang_acceleration_2011,hermsen_onthe_2012,greulich_mutational_2012}, and on evolutionary rescue in structured populations~\cite{Uecker14,Tomasini20,Tomasini22}. 
Specifically, we consider a perfect biostatic drug, which stops division of sensitive bacteria, but we also discuss the generalization of our results to other drug modes of action. When a well-mixed population is subjected to a treatment by a perfect biostatic drug (for a long enough time), the population gets extinct, except if resistant mutants are present when the drug is added \cite{marrec_quantifying_2018,marrec_resist_2020, marrec_adapt_2020}. The population can thus be rescued by resistance. We study the impact of population spatial structure on this process. We also analyze how the time when drug is added impacts the fate of a spatially structured population. 

We show that spatial structure increases the probability that a bacterial population survives treatment by developing resistance through a neutral or deleterious mutation. Specifically, for a given time of addition of the antibiotic, the survival probability of the population increases when the migration rate between demes is decreased, and when the number of demes is increased while keeping the same total population size. We show that this is due to the local fixation of resistant mutants in one or several demes. We study the composition of the population versus time before the addition of drug, and find that it is strongly impacted by spatial structure. We discuss the parameter regimes where spatial structure facilitates rescue by resistance. 
We further study the time needed for resistant bacteria to colonize all demes after drug addition. 
Finally, we show that our main results regarding the impact of spatial structure on population survival to drug extend to more complex spatial structures, and to the case where resistant mutants are present in the inoculum.

\section*{Models and methods}

\begin{figure}[htbp]
    \centering
   \includegraphics[width = 0.8\textwidth]{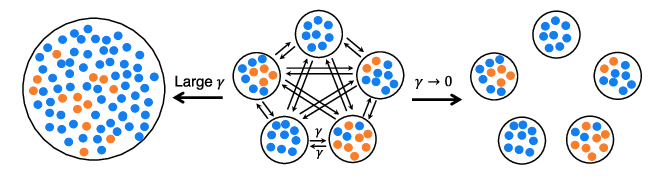}
    \caption{\textbf{Minimal model of spatially structured populations.} We consider a clique structure with one deme per node and a per capita migration rate $\gamma$ (center). For large $\gamma$, it becomes a well-mixed population (left). Meanwhile, if $\gamma \to 0$, it becomes a fully subdivided population. Blue and orange markers represent sensitive (S) and resistant (R) bacteria, respectively.}
    \label{fig:structures-system}
\end{figure}

\paragraph{Spatially structured populations.} We aim to assess the impact of spatial structure on the establishment of antibiotic resistance, in a minimal model where spatial structure is as simple as possible. Thus, we consider a spatially structured bacterial population comprising $D$ demes (i.e.\ subpopulations) on the nodes of a clique (i.e.\ a fully connected graph). This corresponds to the island population model~\cite{Wright31}. Each deme has the same carrying capacity $K$ (Fig \ref{fig:structures-system}, center). For comparison, we also consider a well-mixed population with the same total carrying capacity $DK$ (Fig \ref{fig:structures-system}, left), and a fully subdivided population composed of $D$ demes with carrying capacity $K$ without migrations between them (Fig \ref{fig:structures-system}, right). 
We model migrations from one deme to another through a per capita migration rate $\gamma$, which is the same between all pairs of demes.

We will also briefly discuss extensions to more complex population structures on graphs~\cite{marrec_toward_2021,abbara_frequent_2023}.

\paragraph{Bacterial division, death and resistance.}We consider two types of bacteria: sensitive (S) and resistant (R) ones. In the absence of antibiotics, the two types of bacteria have fitnesses $f_S$ and $f_R$, respectively, which set their maximal division rates (in the exponential phase). They have the same death rate $g$. Thus, we implement selection on birth. However, it is straightforward to generalize our model to selection on death. We assume that growth is logistic in each deme. The division rate of S (resp.\ R) individuals is $f_S \left( 1-N/K\right)$ (resp.\ $f_R \left( 1-N/K\right)$), where $N$ is the current population size of the deme considered. When S bacteria divide, their offspring can mutate to R with probability $\mu$. Here, we focus on the rare mutation regime, where $DK \mu \ll 1$. 

We set $f_S=1$, implying that the time unit in our system is set by the maximum division rate of sensitive bacteria. 
Because antibiotic resistance often carries a fitness cost in the absence of drug~\cite{Andersson99,Andersson10}, we write $f_S=1-\delta$, where $\delta$ represents the cost of resistance and satisfies $0\leq\delta\ll 1$.

We do not model further mutations or backmutations -- note that the rate of backmutations was estimated to be one order of magnitude smaller than the rate of compensatory mutations \cite{levin_compensatory_2000}. 

We consider an ideal biostatic drug that prevents division of the S bacteria, thus changing their fitness to $f_S = 0$. We assume that R bacteria are not affected by the drug. 

\paragraph{Initial conditions and growth regime.} As we aim to model the appearance of resistance, we start with a population comprising only S bacteria. In practice, we initialize each deme (resp. the well-mixed population) with a number of S bacteria representing 10\% of $K$ (resp. of $DK$). Then, each deme (resp.\ the well-mixed population) quickly grows and reaches a steady-state size $N^*=K(1-g/f_S)$ (resp.\ $DN^*=DK(1-g/f_S)$) around which it fluctuates. We choose this initial condition because it is appropriate to model the start of an infection \cite{abel_analysis_2015}. Note that since $DK\mu\ll 1$, it is very unlikely that mutations happen during this quick initial population growth phase, whose timescale is set by $f_S=1$~\cite{marrec_resist_2020}. 
Our results are thus robust to the initialization, as long as it is performed only with S individuals. 

We work in the regime where $K\gg 1$ and $g\ll f_S=1$. Therefore,  extinctions of demes are extremely slow in the absence of drug, and these events can be neglected here, as their timescales are much longer than all those we will discuss. (Specifically, the extinction of a deme with $K=100$ starting with 10 S bacteria, with $g=0.1$ and $f_S=1$, see Ref.~\cite{marrec_adapt_2020}, would take an average time $2.5\times 10^{60}$ in the absence of drug.) Furthermore, in this regime, demes fluctuate weakly around their steady-state size in the absence of drug.

\paragraph{Analytical and numerical methods.} We obtain analytical results from the Moran model \cite{moran_random_1958} and from probability theory. Note that the Moran model assumes that population size is strictly fixed, which is not the case here. However, because we are in the regime where fluctuations around the steady-state size are small in the absence of drug, the Moran process provides good approximations (see Ref.~\cite{marrec_resist_2020}). 

We perform stochastic simulations of the population evolution, using the Gillespie algorithm \cite{gillespie_general_1976, gillespie_exact_1977}. The details of the simulation framework are described in the Supplementary Appendix Section \ref{subs:SI-ssetup}.

\section*{Results}

\subsection*{Spatial structure increases the probability that a bacterial population survives treatment}

How likely is a spatially structured bacterial population to survive biostatic antibiotic treatment? Because our antibiotic prevents S bacteria from dividing while they have a nonzero death rate, a population is doomed to go extinct in the absence of R bacteria. However, if at least one R individual is present when drug is added, \emph{rescue by resistance} can happen \cite{marrec_quantifying_2018, marrec_resist_2020}. Here, we aim to assess the impact of spatial structure on rescue by resistance. To this end, we focus on the probability that the bacterial population survives antimicrobial addition in at least one of the demes. Indeed, R bacteria can then spread to the rest of the population via migrations, ensuring its overall survival.

\begin{figure}[htbp]
        \centering
        \includegraphics[width=\linewidth]{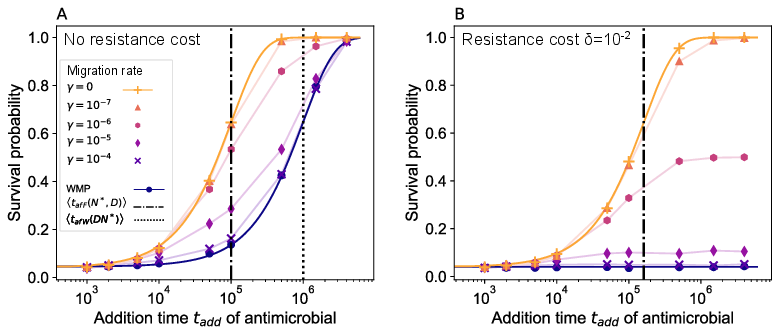}
    \caption{\textbf{Survival probability of a structured bacterial population subjected to biostatic antimicrobial.} The survival probability is shown versus the drug addition time. Panel A: Spatially structured populations composed of $D=10$ demes with $K=100$ are considered, with different per capita migration rates $\gamma$. Results for the well-mixed population with same total size (``WMP") are shown for reference. Markers report simulation results (each obtained from $10^4$ replicate simulations). The vertical dash-dotted line denotes the average time of appearance of a locally successful mutant in the fastest deme (Eq~\ref{eq:fastest}). The vertical dotted line represents the average appearance time of a successful mutant in a well-mixed population of size $DN^*$ (Eq~\ref{eq:tfixgen}). Solid lines for the fully subdivided ($\gamma=0$) and well-mixed populations are analytical predictions from Eq~\ref{eq:psurv-analytical}. Thin solid lines connecting markers are guides for the eye. Panel B: Same as in Panel A, except that mutants have a cost of resistance $\delta =10^{-2}$. Parameter values: $f_S=1$ without drug, $f_S=0$ with drug, $f_R=1$ (Panel A), $f_R=0.99$ (Panel B), $g = 0.1$, $\mu=10^{-5}$.}
    \label{fig:surva-prob}
\end{figure}

Fig \ref{fig:surva-prob}A shows the survival probability versus the time $t_\textrm{add}$ at which antimicrobial is added, for a spatially structured population with different values of the migration rate, in the case where resistant mutants carry no fitness cost, i.e.\ $\delta=0$. At the addition time $t_\textrm{add}$ of antimicrobial, the fitness of S bacteria switches from $f_S=1$ to $f_S=0$, while that of R bacteria remains the same. We observe that survival probability increases with $t_\textrm{add}$. Indeed, R mutants are more likely to appear and fix in the population, which is initially composed only of S bacteria, if there is more time before drug addition. Moreover, we find that survival probability is higher when migration rate $\gamma$ is smaller. In Supplementary Appendix Section \ref{sec:moredemes}, we further observe that subdividing a population with fixed total size into more numerous demes yields higher probabilities of survival. 

Fig \ref{fig:surva-prob}B shows the survival probability versus the drug addition time $t_\textrm{add}$ when resistant mutants have a small cost of resistance $\delta \ll 1$. We observe an even stronger impact of spatial structure on population survival than in the neutral case. Indeed, in the range of $t_\textrm{add}$ considered, the survival probability is roughly constant for the well-mixed population, while it strongly increases with $t_\textrm{add}$ for populations that are fully subdivided or have small migration rates. With the parameter choices of Fig \ref{fig:surva-prob}B, resistant mutants are significantly deleterious in a well-mixed population ($DK\delta \gg 1$), while the impact of the cost is weaker within a deme ($K\delta =1$). Accordingly, the well-mixed population behaves in a very different way in Figs \ref{fig:surva-prob}A and \ref{fig:surva-prob}B, while the fully subdivided one behaves similarly.

Note that the parameter values of Fig \ref{fig:surva-prob} were chosen to be in the rare mutation regime ($DK\mu\ll1$, see ``Models and methods''), and for computational tractability. The parameter regimes where spatial structure facilitates rescue by resistance will be determined in a dedicated section below, and the connection to realistic values will be made in the Discussion.

\subsection*{Local fixation of R mutants promotes the survival of structured populations}

\paragraph{Spatial structure allows local mutant fixation.} To rationalize our results on the impact of spatial structure on the ability of a microbial population to survive the addition of biostatic drug, we focus on rescue by resistance. Indeed, recall that we start from a sensitive inoculum. As mutations occur at division, R mutants arise with rate $N^{*}_\textrm{tot}\mu g$, where $N^{*}_\textrm{tot}$ is the steady-state total number of individuals in the population, $\mu$ the mutation probability upon division and $g$ the death rate, which equals the division rate at steady state. All the populations we considered in Fig \ref{fig:surva-prob} have the same total steady-state size $N^{*}_\textrm{tot}=DK(1-g/f_S)$ in the absence of drug. Thus, the mutant supply is not impacted by spatial structure. However, the fate of mutants can be impacted by spatial structure. 

Most R mutants that appear give rise to lineages that go extinct in the absence of drug, whatever the structure. 
Mutant lineages destined for extinction in the absence of drug can rescue the population if they are present when drug is added~\cite{marrec_resist_2020}. Indeed, once drug is added, S bacteria decay, and R bacteria can proliferate thanks to this reduced competition. However, this scenario remains unlikely in the rare mutation regime ($DK\mu\ll 1$), and is not strongly impacted by the value of $t_\textrm{add}$ or by population structure (see Supplementary Appendix Section \ref{subs:ppres}). Thus, it cannot explain the impact of spatial structure on rescue by resistance that we observe in Fig \ref{fig:surva-prob}. 

Let us now consider mutants that fix. Crucially, in spatially structured populations, fixation can occur locally within a deme. If migrations are rare enough, once R individuals have fixed in a deme, they survive there, provided that migrations are rare enough that we can neglect a subsequent migration and fixation of S bacteria in these R demes. Upon the addition of drug, these locally successful mutants rescue the population. R bacteria can then spread by migration to the whole population.

\paragraph{Timescales of appearance of mutants that locally fix.} To understand the impact of local mutant fixation on population survival, let us focus on successful mutants, i.e.\ those that give rise to a lineage that fixes. The average time $\langle t_{af\, W} (N^*) \rangle$ of appearance (``$a$'') of a mutant that fixes (``$f$'') in a well-mixed population (``$W$'') of fixed size $N^*$ is given by~\cite{desai_beneficial_2007,weissman_rate_2009}:
\begin{equation}
    \langle t_{af\, W} (N^*)\rangle = \langle t_\textrm{app}(N^*)\rangle \times \frac{1}{p_\textrm{fix}(N^*)}=\frac{1}{N^*\mu g\,p_\textrm{fix}(N^*)}\,,
    \label{eq:tfixgen}
\end{equation}
where $\langle t_\textrm{app}(N^*)\rangle=1/(N^*\mu g)$ is the average appearance time of a mutant, and $p_\textrm{fix}(N^*)$ is its fixation probability, which can be expressed in the Moran model~\cite{ewens_mathematical_2004}, see Supplementary Appendix Section~\ref{subs:SI-appwm}. Note that here, we are interested in the time until a successful R mutant appears~\cite{desai_beneficial_2007,weissman_rate_2009}, and not in its fixation time in a deme or in the full structured population~\cite{slatkin_fixation_1981,whitlock2003,hauert_fixation_2014}.

In a structured population, the first mutant that fixes locally in one deme may rescue the population by its resistance. Neglecting the small fluctuations of demes around their steady-state size $N^*=K(1-g/f_S)$, the appearance of such a locally successful mutant can be approximated by a Poisson process with rate $1/\langle t_{af \, W}(N^*)\rangle$ (see Eq~\ref{eq:tfixgen}), and appearance time is exponentially distributed. 
Thus, the average appearance time of a locally successful mutant in the fastest (``$F$'') deme among $D$ demes of steady-state size $N^*$ is (see Supplementary Appendix Section \ref{subs:fastest} for details):
\begin{equation}
    \langle t_{af \, F} (N^*,D)\rangle = \frac{\langle t_{af \, W}(N^*)\rangle}{D}=\frac{1}{DN^*\mu g\,p_\textrm{fix}(N^*)}\,.
    \label{eq:fastest}
\end{equation}

Crucially, a locally successful R mutant appears in the fastest deme of a structured population faster on average than a successful R mutant in a well-mixed population with the same total size. Indeed, Eqs~\ref{eq:tfixgen} and~\ref{eq:fastest} give
\begin{equation}
    \frac{\langle t_{af\, F} (N^*,D)\rangle}{\langle t_{af\, W} (DN^*)\rangle}=\frac{p_\textrm{fix}(DN^*)}{p_\textrm{fix}(N^*)}<1\,,
    \label{eq:ratiotimes}
\end{equation}
for all resistance costs $\delta\geq 0$. This ratio is smaller than 1 because neutral or deleterious mutations fix more easily in small populations than in large ones, due to the increased effect of genetic drift, i.e.\ of fluctuations due to finite-size effects. For neutral R mutants ($\delta=0$), we have $p_\textrm{fix}(N^*)=1/N^*$, so the ratio in Eq~\ref{eq:ratiotimes} is equal to $1/D$. Thus, a locally successful mutant appears in a structured population $D$ times faster than a successful mutant in a well-mixed population with same total size.
For R mutants with a nonzero cost of resistance $\delta>0$, the ratio in Eq~\ref{eq:ratiotimes} is even smaller, as mutant fixation is more strongly suppressed in larger population. See Supplementary Appendix Section \ref{subs:rescost} for details. 

Adding a biostatic antimicrobial (leading to $f_S =0$ and $f_R=1-\delta$ when drug is added) should thus have a different effect depending on the drug addition time $t_\textrm{add}$:
\begin{itemize}
\item If $t_\textrm{add}\ll \langle t_{af \, F} (N^*,D)\rangle$, the bacterial population is likely to be eradicated, whatever its structure.
\item If $ \langle t_{af \, F} (N^*,D)\rangle\ll t_\textrm{add} \ll \langle t_{af \, W} (DN^*)\rangle$, it is likely that a locally successful mutant has appeared. This rescues a fully subdivided population (with no migrations). However, it is likely that no successful mutant has appeared in a well-mixed population yet. A well-mixed population is thus expected to go extinct. 
\item If $t_\textrm{add} \gg \langle t_{af \, W} (DN^*)\rangle$, both a well-mixed and a fully subdivided population are expected to survive.
\end{itemize}

In Fig \ref{fig:surva-prob}A, where neutral mutants are considered, we show the two timescales compared in Eq~\ref{eq:ratiotimes}. We observe that
the most substantial impact of structure on survival probabilities is observed when $\langle t_{af\, F}(N^*,D)\rangle < t_\textrm{add} < \langle t_{af\, W} (DN^*)\rangle$. This is fully in line with our theoretical analysis based on timescale comparisons. Furthermore, we observe that all populations survive if $t_\textrm{add} \gg \langle t_{af \, W} (DN^*)\rangle$. Finally, if $t_\textrm{add}\ll \langle t_{af \, F}(N^*,D) \rangle$, the population is eradicated in most cases. It can nevertheless be rescued by mutant lineages that were destined for extinction in the absence of drug. Thus, the probability of treatment survival is then given by the probability that non-successful mutants are present when the drug is added, see Eq~\ref{eq:ppres}. 

In Fig \ref{fig:surva-prob}B, we consider mutants with moderate fitness cost $\delta=1/K$. The average appearance time $\langle t_{af\, F} (N^*,D)\rangle$ of a successful mutant in the fastest deme is shown (cf. Eq~\ref{eq:fastest} and Eq~\ref{eq:tapp-fastest-cost}) and is of the same order of magnitude as in the neutral case of Fig \ref{fig:surva-prob}A.  By contrast, these mutants are significantly deleterious in the well-mixed population since $DK\delta=D\gg 1$. Hence, $\langle t_{af\, W} (DN^*) \rangle$ is not shown in Fig \ref{fig:surva-prob}B, as it evaluates to $2.2\times 10^9$, which is about $10^4$ times longer than $\langle t_{af\, F} (N^*,D)\rangle$. The impact of spatial structure is even stronger in this case than for neutral mutants. 

\paragraph{Analytical expression of the survival probability.} To go beyond our comparison of timescales, let us express the probability $p_\textrm{surv} (t_\textrm{add})$ that a population survives the addition of drug at time $t_\textrm{add}$, in a well-mixed population or in a fully subdivided population with no migrations. For this, we approximate deme size as constant, and we consider the two distinct mechanisms that can yield population survival, namely rescue by a (locally) successful mutant and rescue by the presence of the lineage of a mutant that was destined for extinction in the absence of drug. This leads to:
\begin{equation}
    p_\textrm{surv} (t_\textrm{add}) = p_\textrm{succ}(t_\textrm{add}) + \left[1-p_\textrm{succ}(t_\textrm{add})\right] p_\textrm{pres}\,,
    \label{eq:psurv-analytical}
\end{equation}
where $p_\textrm{succ}(t_\textrm{add})$ is the probability that a (locally) successful mutant has appeared by $t_\textrm{add}$, while $p_\textrm{pres}$ is the probability of presence of a lineage of R mutants destined to go extinct in the absence of drug. In the Supplementary Appendix, we calculate $p_\textrm{pres}$ in Section \ref{subs:ppres} (see Eq~\ref{eq:ppres}) and $p_\textrm{succ}(t_\textrm{add})$ in Section \ref{sec:purv_analytical}. The specific form of $p_\textrm{succ}(t_\textrm{add})$ is given by Eqs~\ref{eq:psucc-W-nocost} and \ref{eq:psucc-W-cost} for a well-mixed population, and by Eqs~\ref{eq:psucc-S-nocost} and \ref{eq:pucc-S-cost} for a fully subdivided population (with and without fitness cost, respectively). 

Fig \ref{fig:surva-prob} shows that our simulation results are in very good agreement with the analytical predictions from Eq~\ref{eq:psurv-analytical}, both for the well-mixed population and for the fully subdivided population. 

\subsection*{Spatial structure impacts population composition before drug addition}
To understand in more detail the role of spatial structure on resistance spread, let us study the system composition before drug is added. For simplicity, let us perform this analysis in the case of neutral R mutants, with the same parameter values as in Fig \ref{fig:surva-prob}A.

\begin{figure}[h!]
    \centering
    \includegraphics[width=0.92\textwidth]{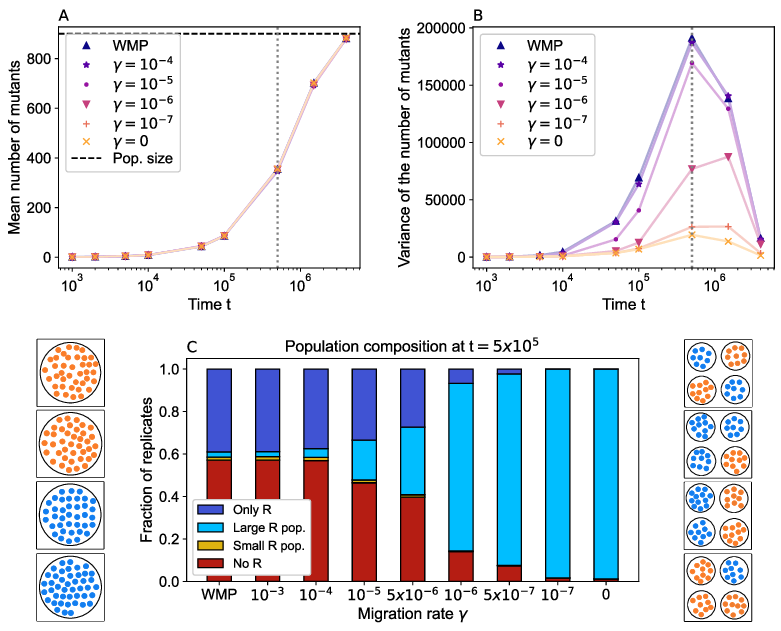}
    \caption{\textbf{Impact of spatial structure on population composition in the absence of drug.} 
    Panel A: The mean across replicates of the total number of R mutants in the population is shown versus time for population structures differing by the migration rate $\gamma$, and for the well-mixed population with same total size (``WMP"). Horizontal dashed line: steady-state total population size $K\left( 1- g/f_R \right)$. Panel B: Variance across replicates in the total number of R mutants in the population, in the same conditions as in panel A. Vertical dotted line in panels A and B: time when largest variance was obtained in panel B ($t=5\times10^5$). Panel C: Population composition at $t=5\times10^5$, for different migration rates. We distinguish four cases. ``Only R'': R bacteria have fixed in the whole population; ``Large R population'': at least one deme comprises more than 9 R individuals, but the R type has not fixed in the whole population; ``Small R population'': at least one deme comprises between 1 and 9 R bacteria, but no deme has more than 9 R; ``No R'': there are no R bacteria. We distinguish ``Large'' and ``Small R population'' to account for the possibility of stochastic extinction of R lineages, which becomes negligible if $\gtrsim 10$ R bacteria are present in a deme (see Supplementary Appendix Section~\ref{subs:stochext}). Results are obtained from $10^{4}$ replicate simulations. Parameter values (in all simulations): $K=100$, $D=10$, $f_S=f_R=1$, $g = 0.1$, $\mu=10^{-5}$. Left and right of panel C: Schematics showing typical population states in four replicates (square boxes) for a well-mixed (left) and a fully subdivided population composed of $D=4$ demes (right). Blue: S bacteria; orange: R bacteria. } 
    \label{fig:composition-T5e5}
\end{figure}

Fig \ref{fig:composition-T5e5}A shows the time evolution of the total number of R mutants present in a structured population, averaged over many stochastic simulation replicates. 
Whatever the migration rate, the average number of R individuals grows as mutants fix in some simulation replicates. This occurs either first locally for structured populations, or directly in the whole population for the well-mixed population (see Supplementary Appendix Section \ref{subs:growth}). Despite this difference, the average over replicates is not impacted by population structure. Indeed, for neutral mutants, using Eq~\ref{eq:tfixgen} and $p_\textrm{fix}(N^*)=1/N^*$ gives an average time $\langle t_{af \, W}(N^*) \rangle=1/(\mu g)$ for a successful mutant to appear in a well-mixed population. As this timescale is independent of population size, it governs the process of R mutants taking over each single deme but also a larger well-mixed population. 

Fig \ref{fig:composition-T5e5}B shows that the variance across replicates of the total number of R mutants in a population is strongly impacted by its structure. This is particularly true in the time interval between $\langle t_{af\, F} (N^*,D)\rangle$ and $\langle t_{af\, W} (DN^*)\rangle$ (see Fig \ref{fig:surva-prob}A). This suggests that the impact of population structure on the variance of mutant number is driven by R mutant fixation. Fig~\ref{fig:composition-T5e5}B further shows that the variance of the number of mutants across replicates is larger when migrations are more frequent. In a well-mixed population, mutants fix in one step, while in a fully subdivided population, they fix separately in each deme (see Supplementary Appendix Section \ref{subs:growth}). The large variance observed for the well-mixed population comes from the variability across replicates of the appearance time of a successful mutant. For the fully subdivided population, this is smoothed by lumping together $D$ separate demes. Within each deme, the variance across replicates of the number of R mutants is not impacted by spatial structure, if R mutants are neutral, see Supplementary Appendix Section~\ref{subs:dyndeme}.

In Fig \ref{fig:composition-T5e5}C, we examine in more detail the population composition in the absence of drug, at the time when we obtained the largest variance in Fig \ref{fig:composition-T5e5}B (vertical dotted line). We observe that in most replicates, the well-mixed population has either fixed resistance or does not have any R mutants. Thus, the number of mutants is either 0 or $(1-g/f_R)DK$, yielding the large variance seen in Fig \ref{fig:composition-T5e5}B.  Typical compositions of the well-mixed population are illustrated schematically on the left side of Fig \ref{fig:composition-T5e5}C. As we partition the system spatially and reduce the migration rate, structured populations more frequently include some fully R demes and some fully S ones. Consequently, the fraction of replicates showing overall fixation or overall extinction decreases. This results in a smaller variance across replicates, as the number of mutants becomes more homogeneous across them. Typical compositions of the fully subdivided population are illustrated schematically on the right side of Fig \ref{fig:composition-T5e5}C. 

\begin{figure}[htbp]
    \centering
    \includegraphics[width = 0.92\textwidth]{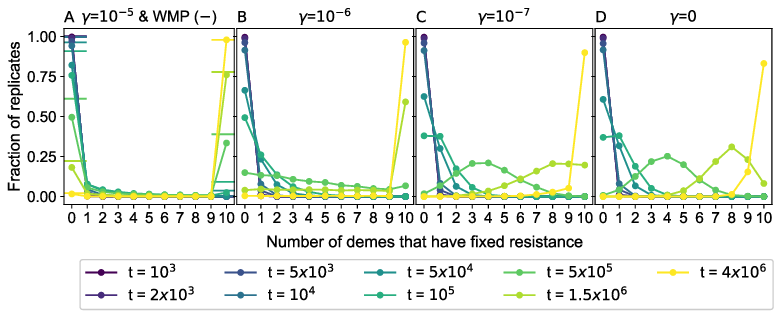}
    \caption{\textbf{Number of demes that have fixed resistance before drug application.} Histograms of the number of demes that have fixed resistance, computed across $10^4$ simulation replicates, are shown for different values of the elapsed time $t$ (different colors), and of the migration rate $\gamma$ (different panels). Specifically, for each $i\in [0, 10]$, we show the fraction of realizations in which $i$ demes have fixed resistance. The horizontal line markers in panel A show the results for a well-mixed population with same total size (``WMP"). Parameter values: $K=100$, $D=10$, $f_S=f_R=1$, $g = 0.1$, $\mu=10^{-5}$.}
    \label{fig:fixR}
\end{figure}

An important cause of the composition difference between structured and well-mixed populations is the possibility of local fixation of R mutants.
What is the fixation dynamics at the deme level before we add the drug? In Fig \ref{fig:fixR}, we report how many demes have fixed resistance for different migration rates and at different times, in the form of histograms computed over simulation replicates. Note that for the well-mixed population we report overall fixation. As expected, the number of demes where R mutants have fixed increases over time. Furthermore, we observe that the distribution of demes that have fixed resistance and its dynamics are strongly impacted by spatial structure. For large migration rates (panel A), at all times, most replicates have fixed resistance in either no deme or all demes. This is close to the large-$\gamma$ limit (i.e.\ the well-mixed population), shown as horizontal lines in panel A. When the migration rate $\gamma$ is decreased (panels B and C), the fraction of realizations featuring an intermediate number of demes that have fixed resistance increases at intermediate times. For the small migration rate $\gamma=10^{-7}$ (panel C), the distributions become close to those observed for $\gamma = 0$ (panel D). Thus, small migration rates result in a population with a transient strong heterogeneity across demes. This has a crucial impact on the outcome of drug treatment, and on the survival probability curves in Fig \ref{fig:surva-prob}. Indeed, the biostatic drug does not affect R mutants, and the presence of at least one deme where R has fixed rescues the population when the drug is applied. Population composition is further shown versus time for different migration rates in the Supplementary Appendix Section~\ref{subs:compodeme}.

\subsection*{Spatial structure facilitates rescue by resistance in broad parameter regimes}

\paragraph{Mutation rate and population size.} As mentioned in ``Models and methods'', we focus on the rare mutation regime $DK\mu\ll 1$. Outside of this regime, R mutants will usually be present in the population, and allow its rescue, independently of spatial structure. Another point where population size could have a role is that it takes longer for a lineage of R mutants destined to fix in a deme to actually fix there if the deme is larger. However, this does not affect our rescue scenario, because as long as the number of R mutants that are present when drug is added is sufficient for stochastic extinction not to occur after the addition of drug, the population will be rescued. 

\paragraph{Cost of resistance.} For neutral resistant mutants ($\delta=0$), we saw that a locally successful mutant appears in a structured population $D$ times faster than a successful mutant in a well-mixed population with same total steady-state size $DN^*$, where $N^*$ is the steady-state size of a deme. This leads to a substantial effect of structure on evolutionary rescue by drug resistance, see Fig~\ref{fig:surva-prob}A.
For mutants with a substantial cost of resistance $\delta\gg 1/K$, the mutant fixation probability in a deme is $p_\textrm{fix}(N^*)\approx \delta e^{-N^*\delta}$ (see Supplementary Appendix Section \ref{subs:SI-appwm}), and thus the time until a successful mutant appears in a well-mixed population grows exponentially with $N^*$ (see Eq~\ref{eq:tfixgen}). The effect of spatial structure on rescue by resistance is then even stronger than for neutral mutants, but it exists at exponentially longer timescales (see Supplementary Appendix Section \ref{subs:rescost}). Finally, mutants with intermediate cost satisfying $1/(DK)\ll\delta\ll1/K$ are effectively neutral in demes of size $N^*$, but substantially deleterious in a well-mixed population of size $DN^*$. This regime yields a strong effect of spatial structure, which extends from the same timescales as in the neutral case to longer ones, as in Fig~\ref{fig:surva-prob}B (note that mutants are slightly more costly there: $\delta=1/K$). Thus, spatial structure facilitates evolutionary rescue by resistance for all costs $\delta\geq 0$, but the effect exists for extremely long timescales if $\delta\gg 1/K$. Because of this, we focus on small costs.

\paragraph{Migration rate.} So far, we mainly considered the extreme cases of the fully subdivided population and of the well-mixed population for our analytical reasoning. With the parameter values chosen in Fig \ref{fig:surva-prob}, we observe that in practice, the transition between these two extreme cases occurs when the per capita migration rate is increased from $\gamma=10^{-7}$ to $\gamma=10^{-4}$. More generally, what is the relevant range of migration rates where spatial structure facilitates evolutionary rescue by drug resistance? Given the discussion above on the cost of resistance, let us focus on R mutants with cost satisfying $\delta \ll 1/K$. These mutants are neutral or effectively neutral within demes, and the fixation probability of an R (resp.\ S) individual in a deme of S (resp.\ R) bacteria is $p_\textrm{fix}(N^*)=1/N^*$. 

We determined that spatial structure can facilitate evolutionary rescue by drug resistance because it fosters local fixation of R mutants. The demes that fix R mutants act as refugia for R mutants, and allow population survival when drug is added. For this to happen, local fixation of R mutants must be possible, without being perturbed by S bacteria migrating from other demes during the fixation process. The average fixation time of a neutral mutant in a well-mixed population of size $N^*$ is $\langle t_\textrm{fix}\rangle=(N^*-1)/g$, where we used the Moran process result~\cite{ewens_mathematical_2004}. Meanwhile, the expected time until an S individual with a lineage destined for local fixation migrates into the deme where R is in the process of taking over is $\langle t_\textrm{mig}\rangle=1/[N^*\gamma(D-1)p_\textrm{fix}(N^*)]=1/[\gamma(D-1)]$. Thus, for incoming migrations not to strongly perturb local fixations, we need $\langle t_\textrm{fix}\rangle\ll\langle t_\textrm{mig}\rangle$, i.e.
\begin{equation}
    \frac{\gamma}{g}\ll\frac{1}{(D-1)(N^*-1)}\,.
    \label{eq:ratherraremigr}
\end{equation}
Accordingly, in Fig \ref{fig:surva-prob}A, spatial structure significantly impacts population survival when Eq~\ref{eq:ratherraremigr} holds, i.e.\ when $\gamma\ll 10^{-4}$, and results become similar to those of a well-mixed population when $\gamma=10^{-4}$. Note that Eq~\ref{eq:ratherraremigr} is less stringent than the rare migration regime~\cite{slatkin_fixation_1981,hauert_fixation_2014,marrec_toward_2021}, which requests a separation of timescales between migration and fixation processes. Here, we took into account the fact that only some migration events may strongly impact deme composition.

Once the fastest deme has fixed resistance, S bacteria may still migrate to that deme and fix there, leading to the extinction of R mutants. The expected time until this happens is again $\langle t_\textrm{mig}\rangle=1/[\gamma(D-1)]$. Let us compare this to the average appearance time of a mutant that fixes in the fastest deme, given in Eq~\ref{eq:fastest}, i.e.\ $\langle t_{af \, F} (N^*,D)\rangle=1/(D\mu g)$ for effectively neutral mutants. If $\langle t_\textrm{mig}\rangle\gg \langle t_{af \, F} (N^*,D)\rangle$, R mutants that locally fixed are expected to survive until other locally successful mutants appear. Then, provided that $t_\textrm{add}\gg \langle t_{af \, F} (N^*,D)\rangle$, R mutants are present when drug is added, leading to rescue by resistance. Thus, spatial structure is expected to facilitate evolutionary rescue by resistance about as much as in a fully subdivided population if $\langle t_\textrm{mig}\rangle\gg \langle t_{af \, F} (N^*,D)\rangle$, i.e.
\begin{equation}
    \frac{\gamma}{\mu g}\ll \frac{D}{D-1}\approx 1\,.
    \label{eq:veryraremigr}
\end{equation}
Indeed, in Fig \ref{fig:surva-prob}A, structured populations with migrations and fully subdivided ones have very similar survival probabilities when $\gamma\ll 10^{-6}$, i.e.\ when Eq~\ref{eq:veryraremigr} is satisfied. Survival probabilities intermediate between those of well-mixed populations and of fully subdivided populations are observed when Eq~\ref{eq:ratherraremigr} is satisfied but not Eq~\ref{eq:veryraremigr}. Note that while $\langle t_\textrm{mig}\rangle$ is the expected time until the next S individual destined for fixation migrates into the single R deme, R individuals from that deme may migrate to other demes. Taking this into account gives an expected extinction time of R mutants which is $\langle t_\textrm{mig}\rangle$ times a prefactor of order $\log(D)$, assuming rare migrations~\cite{bitbol_quantifying_2014}.

\subsection*{R mutants readily colonize the whole population after drug addition}

So far, we focused on the impact of spatial structure on the survival probability of the population upon drug addition. Let us now ask what happens after drug addition.
With biostatic drug, S bacteria cannot divide or mutate, and their decay leaves some demes empty. 
If R mutants rescue a spatially structured population because they were present in at least one deme before drug addition, how fast do they then spread in all demes? This question has implications for disease recurrence, where after an initial transient alleviation of symptoms due to the drug-induced decay of S bacteria, the spread of R mutants could lead to a regrowth of the pathogenic bacteria population.

\begin{figure}[htb!]
        \centering
    \includegraphics[width=0.92\textwidth]{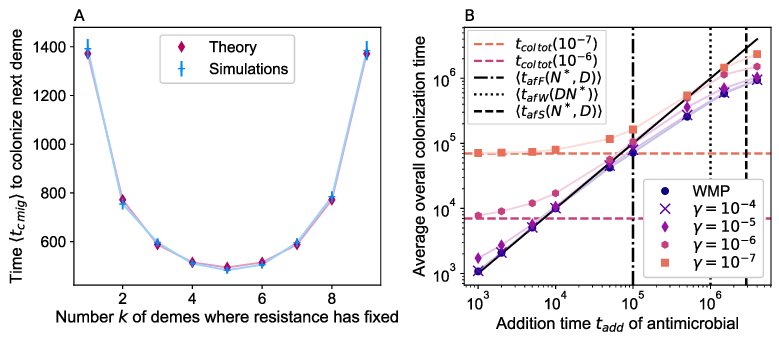}
\caption{\textbf{Colonization of the whole population by R mutants}. Panel A: The time $\langle t_\textrm{c mig} \rangle$ to colonization of the next deme by R mutants through migrations after drug addition is shown versus the number $k$ of demes where R mutants have already fixed. Migration rate: $\gamma=10^{-6}$. We compare simulation results from $5\times 10^3$ realizations with the theoretical prediction from Eq~\ref{eq:taufmig} (``Theory''). Error bars represent 95\% CI. Panel B: The average time to complete colonization by R mutants (i.e., when the number of R mutants first becomes at least equal to $0.9 K$ in each separate deme) is shown versus the drug addition time $t_\textrm{add}$ for structured populations with different migration rates $\gamma$, and for the well-mixed population with same total size (``WMP"). Out of $10^4$ replicate simulations, we restrict to those where the population survived treatment.  
Horizontal lines: analytical prediction for $\langle t_\textrm{c tot}\rangle$ from Eq~\ref{eq:Tftot} for the two lowest $\gamma$ values. Vertical lines: average local fixation time of R mutants in the fastest deme ($\langle t_{af\, F}(N^*,D) \rangle$), in a well-mixed population ($\langle t_{af\, W}(DN^*)\rangle$), and in the slowest deme ($\langle t_{af\, S} (N^*,D)\rangle$). The black line $x=y$ separates colonization occurring before drug is added (below the line) and after (above it). Confidence intervals are too small to be visible in this logarithmic scale. Parameter values for both panels: $K=100$, $D=10$, $f_S=f_R=1$, $g = 0.1$, $\mu=10^{-5}$.}
    \label{fig:timescales}
\end{figure}

To address this question, let us first calculate the time $\langle t_\textrm{c mig} \rangle$ it takes for an R mutant to colonize an empty deme, starting from a population where resistance has fixed in $k\geq 1$ demes out of $D$, while other demes are empty. Then, there are $ N^* k $ mutant individuals, where $N^*$ is the steady-state deme size, which can migrate to each of the $D-k$ empty demes at a per capita rate $\gamma$. However, once an R mutant arrives in an empty deme, its lineage may go stochastically extinct. This happens with probability $1-g/f_R$ (see Supplementary Appendix Section \ref{sec:corr-stochext}). 
Thus, we have:
\begin{equation}
\langle t_\textrm{c mig} (k) \rangle = \frac{1}{\gamma N^* k (D-k)(1-g/f_R)}\,.
\label{eq:taufmig}
\end{equation}
In Fig \ref{fig:timescales}A, we compare the analytical prediction in Eq~\ref{eq:taufmig} to our simulation results, obtaining a good agreement. The time $\langle t_\textrm{c mig} \rangle$ for R mutants to colonize the next deme features a minimum when half of the demes are mutant. Indeed, more mutant demes yield more mutants that may migrate, but fewer wild-type demes where they can fix, leading to a trade-off.

Eq~\ref{eq:taufmig} neglects the presence of S bacteria in the population, which would affect the stochastic extinction of R lineages. This is acceptable provided that the extinction of S bacteria upon drug addition is fast enough.
The decay time upon drug addition of a well-mixed population comprising $N^*$ S bacteria is $\tau_S = \left(1/g\right)\sum_{i=1}^{N^*} 1/i$~\cite{marrec_resist_2020}. For the migration rate considered in Fig \ref{fig:timescales}A, $\tau_S$ is indeed negligible with respect to $\langle t_\textrm{c mig} \rangle$ (see Supplementary Appendix Section~\ref{sec:colSI}). More generally, the condition $\tau_S\ll\langle t_\textrm{c mig} (1) \rangle$ yields $\gamma/g\ll 1/[N^*(D-1)\sum_{i=1}^{N^*}1/i]$, which is holds in most cases where spatial structure favors rescue by resistance (see Eq~\ref{eq:ratherraremigr}). The condition $\tau_S\ll\langle t_\textrm{c mig} (N^*/2) \rangle$ is however slightly more demanding, as it yields $\gamma/g\ll4/[N^*D^2\sum_{i=1}^{N^*}1/i]$.

So far, we focused on one step of the spread of R mutants. How long does it take in practice for R mutants to colonize the whole structured population after drug addition? First, if drug is added after a short time $t_\textrm{add}<\langle t_{af\,F}(N^*,D)\rangle$, at most one deme in the population should comprise R mutants or have fixed them. Then, the average time $\langle t_\textrm{c tot}\rangle$ it takes for R mutants to colonize the whole population can be obtained by summing the step-by-step colonization times in Eq~\ref{eq:taufmig}, see Supplementary Appendix Section~\ref{sec:colSI}. 
In Fig \ref{fig:timescales}B, we report the overall colonization time of the population by R mutants, defined as the time $t_\textrm{add}$ when drug is added plus the time needed for colonization after drug addition. We observe that for small addition times satisfying both $t_\textrm{add} \ll \langle t_{af\, F}(N^*,D)\rangle$ and $t_\textrm{add} \ll \langle t_\textrm{c tot}\rangle$, the colonization time is indeed well-described by $\langle t_\textrm{c tot}\rangle$. 
It leads to a plateau whose value is governed by the migration rate. 
Note that in Fig \ref{fig:timescales}B, this plateau is only observed for small migration rates ($ \gamma \lesssim 10^{-5}$) such that the condition $t_\textrm{add} \ll \langle t_\textrm{c tot}\rangle$ is satisfied for the smallest $t_\textrm{add}$ values considered. 
For larger values of $t_\textrm{add}$ and larger migration rates, the timescale $t_\textrm{add}$ dominates the overall colonization time of the population by R mutants, as $t_\textrm{add}>\langle t_\textrm{c tot}\rangle$, see Fig \ref{fig:timescales}B. In these cases, mutant spread can be considered fast when drug is added. Finally, if $t_\textrm{add}$ is further increased, R mutants may fix in the whole population (i.e.\ colonize it) before drug addition. Thus, the overall colonization time becomes smaller than $t_\textrm{add}$ and tends toward a new plateau. 
Note that in this regime, colonization can occur before drug addition through migrations or new mutations, followed by fixations. The time it takes for a locally successful mutant to appear in the slowest deme can be expressed as $\langle t_{af\, S} (N^*,D)\rangle = \langle t_{af \, W} (N^*)\rangle \sum_{i=1}^D 1/i$ (see Supplementary Appendix Section \ref{subs:slowestfix}). It is beyond this time that we expect a plateau even for small migration rates. This is indeed what is observed in Fig \ref{fig:timescales}B.

\subsection*{Extensions to more complex spatial structures and heterogeneous inoculum}

\paragraph*{Rescue by resistance is also facilitated by spatial structures beyond the clique.}
So far, we considered a minimal model of spatially structured population, known as the clique or the island model, where all demes are equivalent and connected to one another by identical migration rates (see ``Models and methods''). We showed that spatial structure facilitates evolutionary rescue through the local fixation of resistant mutants in demes. This mechanism can also exist in different spatial structures, where demes are placed on graphs other than a clique.  

We explore the impact of spatial structure on population survival upon drug addition in three structures with more reduced symmetry than the clique in the Supplementary Appendix Section \ref{sec:diff-popstruct}. First, we consider a two-dimensional square lattice with periodic boundary conditions. 
Second, we consider a star comprising a central deme connected to $D-1$ leaves, where all leaves are equivalent~\cite{marrec_toward_2021}. 
Third, we consider a line (or linear stepping-stone model) composed of $D$ demes, where each deme is connected to two neighbors, except at the two ends~\cite{servajean_impact_2024}. In all cases, we choose the migration rates so that the total incoming migration rate to a deme is the same as in the clique, with the exception of the central deme in the star with asymmetric migrations, and of the end demes in the line with asymmetric migrations. This makes the re-invasion of an R deme by S bacteria as similar as possible across structures, but not fully.
Our simulation results in the Supplementary Appendix Section \ref{sec:diff-popstruct} show that all these structured populations have the same probability to survive treatment if $t_\textrm{add}<\langle t_{af \, F}(N^*,D) \rangle$, i.e., in cases where usually not more than one deme has fixed resistance before drug addition. For larger values of drug addition time $t_\textrm{add}$, our results for the star depend on migration asymmetry and significantly differ from those obtained for other structures, although the difference remains minor. We interpret this difference as arising from S individuals possibly re-invading demes where R mutants have fixed, which remains structure-dependent. 

Overall, our main result that spatial structure increases survival probability upon drug addition is qualitatively robust to changing the graph on which demes are placed, as it largely depends on local fixation of R mutants.
However, migration patterns impact the re-invasion of R demes by S bacteria, and because of this, the conditions on migration rate we identified in Eqs~\ref{eq:ratherraremigr} and~\ref{eq:veryraremigr} for the clique will differ for other graphs. The spread of R mutants to other demes by migration before drug addition, and the colonization of the graph by R mutants after drug is added, will also be affected by graph structure.

\paragraph*{Inoculum that contains R mutants.}
So far, we focused on a sensitive inoculum and on resistant individuals appearing through mutations (see ``Models and methods''). However, resistant mutants can be present in the inoculum. This can be promoted e.g.\ by the presence of antibiotic in other hosts or in the environment. Furthermore, spatially structured populations starting with a fraction of R mutants in each deme were recently considered in \textit{in vitro} experiments and modeling in Ref.~\cite{kreger_role_2023}. To connect to these cases, we consider the initial condition where some mutants are already present in each deme of a clique, see Supplementary Appendix Section~\ref{mutino}. 
We observe that in this case too, spatial structure increases the survival probability of the population. Indeed, mutant extinction in all demes is less likely than mutant extinction in a well-mixed population.

\section*{Discussion}
Here, we showed that spatial structure can facilitate the survival of a bacterial population to antibiotic treatment, starting from a sensitive inoculum. 
Indeed, the bacterial population can be rescued if antibiotic resistant mutants are present when drug is added. While the emergence of resistant bacteria by random mutations only depends on total population size and not on spatial structure, their fate can be affected by spatial structure. If the mutation that provides resistance is neutral or deleterious, which is usually the case, its probability of fixation is increased in smaller populations. This leads to local fixation of resistant mutants in demes, which then constitute refugia of resistant mutants and allow the population to survive when drug is added. Because of this, spatial structure facilitates evolutionary rescue by drug resistance. The survival probability of the population increases when the migration rate between demes is decreased, and when the number of demes is increased. We showed that this effect exists in the rare mutation regime, for relatively rare to rare migration rates, and we quantified these conditions. While it holds for all costs of resistance, the timescales involved become very long for substantial costs. Thus, we mainly focused on neutral resistant mutants and those with moderate costs. After drug is added and the population is rescued by resistance, migrations allow resistant mutants to colonize all demes. While we considered a minimal model where all demes are equivalent and connected to each other with identical migration rates, bacterial populations can have diverse spatial structures. We showed that our key finding that spatial structure facilitates rescue by resistance still holds for more complex spatial structures, and when resistant mutants are present in the inoculum, as it relies on local fixation.

The effect of spatial structure we evidenced here is due to stochastic fixation of neutral mutants. Stochastic effects often give rise to original behaviors of small and structured populations. In particular, inoculum size impacts the survival of bacterial population in the presence of antibiotic~\cite{Coates18,alexander_stochastic_2020}, and spatial partitioning impacts the efficiency of antibiotic resistance by production of the beta-lactamase enzyme \cite{verdon_spatial_2024}. However, the effect evidenced here is different, as it relies on the stochasticity in the fate of a mutant, and not in the stochasticity in the inoculum size. 

Because local fixation of neutral resistant mutants is instrumental to our effect, it involves rather long timescales. For neutral resistant mutants, the average appearance time $\langle t_{af \, F} \rangle = 1/(D\mu g)$ of a locally successful mutant in the fastest deme is of particular relevance. As it scales as the inverse mutation probability per individual and per generation $1/\mu$, it can be quite long, even considering that there are often several different mutational targets that give rise to resistance, leading to a larger effective $\mu$. However, this timescale is inversely proportional to the number $D$ of demes, meaning that it can potentially become arbitrarily small when increasing population subdivision. Recall however that we need to be in the rare mutation regime for our effect to be important. With a mutation probability per nucleotide and per generation of $10^{-10}$~\cite{lee_rate_2012}, assuming 10 possible mutations for the development of resistance to a given drug, gives $\mu=10^{-9}$. Then, for bacteria dividing once per hour, we find $\langle t_{af \, F}\rangle=10^8$~h for $D=10$, but $\langle t_{af \, F}\rangle=10^3$~h for $D=10^6$. In the latter case, we would be in the rare mutation regime for deme sizes up to a few hundred bacteria.

Host-associated bacterial populations are often fragmented into numerous small demes, for instance in intestinal crypts, lymph nodes, or skin pores. First, the gut features a rich microbiota which can be a reservoir of resistant bacteria in humans and animals. Intestinal crypts are glandular structures located at the base of the gut lining. The mouse intestine comprises approximately $D=10^5$ crypts~\cite{casteleyn_surface_2010}, each harboring about 100 to 400 bacteria~\cite{pedron_acrypt_2012,dekaney_expansion_2007}, and the human intestine comprises up to $D=10^7$ crypts~\cite{Nguyen10}. Second, lymph nodes may act as reservoirs of bacteria, especially as some antibiotics poorly permeate them~\cite{ganchua_lymph_2020}. A mouse possesses about $D=20$ lymph nodes~\cite{van2006anatomy}, and each mesenteric lymph node holds $10^3$ to $10^5$ bacteria~\cite{hapfelmeister_thesalmonella_2005}. Finally, a human face features roughly $D=\SI{2e4}{}$ skin pores~\cite{campos_useof_2019}, and a typical skin pore contains about \SI{5e4}{} \emph{Cutibacterium acnes} bacteria~\cite{conwill_anatomy_2022, claesen_antibiotic_2020}. These orders of magnitude (detailed in Supplementary Appendix Section \ref{sec:bioloex}) show that the effect of spatial structure on rescue by resistance that we evidenced here may be relevant in such real-world structured populations.

In this work, we focused on a perfect biostatic drug that prevents any division of sensitive bacteria. 
However, our main findings generalize to biocidal drugs that increase the death rate of sensitive bacteria, or to drugs which combine both modes of actions. Indeed, what happens before the addition of drug, in particular the local fixation of resistant mutants, is not impacted by the mode of action of the drug. Thus, the effect of spatial structure we evidenced here holds independently of this. The difference between the biostatic and the biocidal case lies in the decay of the sensitive bacteria once drug is added. In the biocidal case, sensitive microbes may still divide during this phase, and new resistant mutants may then appear~\cite{marrec_resist_2020,czuppon_stochastic_2023}. This is the case both in well-mixed and in structured populations, with only minor differences between them, due to the stochasticity of extinction. Thus, our main findings are robust to the mode of action of the drug. 
Beyond modes of action, it would be interesting to study the impact of spatial structure on multi-step drug resistance evolution~\cite{Igler21}.

Here, we considered a minimal model of spatial structure, without any environmental heterogeneities. This allowed us to find a generic effect of spatial structure on the survival of a population to antibiotic treatment. It would be very interesting to extend our work to heterogeneous environments, which are known to have a substantial effect on antibiotic resistance evolution~\cite{zhang_acceleration_2011,hermsen_onthe_2012,greulich_mutational_2012}, and on evolutionary rescue in structured populations~\cite{Uecker14,Tomasini20,Tomasini22}.

Some mechanisms of resistance to antibiotics involve the production of a public good. One prominent example is the beta-lactamase enzyme, which degrades beta-lactam antibiotics in the environment. Spatial structure was recently experimentally shown to have an important impact in this case~\cite{verdon_spatial_2024}. Note that collective protection via a public good exists in multiple other cases~\cite{yurtsev_oscillatory_2016, domingues_social_2017, galdino_sideropores_2024}. Recently, a model was developed to describe this phenomenon in a well-mixed population~\cite{hernandez_coupled_2023, hernandez_ecoevolutionary_2024}. Extending this study to the case of structured populations would be very interesting. More generally, coupling these evolutionary questions to ecological interactions, which also have an interesting interplay with spatial structure~\cite{limdi_asymmetric_2018, gokhale_migration_2018,wu_modulation_2022}, is an exciting perspective.

It would be interesting to test our predictions experimentally. \emph{In vitro} experiments considering bacterial populations with migrations between demes have been performed \cite{Kryazhimskiy12,Nahum15,france_relationship_2019,chakraborty_experimental_2023,kreger_role_2023}, but generally with a constant environment. Investigating the fate of populations upon the addition of drug in these setups would be very interesting. A challenge is that the timescales relevant here are long. However, they can become smaller if the number $D$ of demes is increased. Long experiments with large numbers of demes can be achieved using robots for serial passage~\cite{Kryazhimskiy12,Horinouchi14,Maeda20}, and the connection between theory and experiment for spatially structured populations is progressing~\cite{abbaraPreprint}.

\section*{Code availability} 
Python code for our stochastic simulations is freely available at \url{https://github.com/Bitbol-Lab/DrugRes_StructPop}

\section*{Acknowledgments}
This research was partly funded by the Swiss National Science Foundation (SNSF) (grant No.~315230\_ 208196, to A.-F.~B.) and by the European Research Council (ERC) under the European Union’s Horizon 2020 research and innovation programme (grant agreement No.~851173, to A.-F.~B.).


\renewcommand{\thefigure}{S\arabic{figure}}
\setcounter{figure}{0}
\renewcommand{\thetable}{S\arabic{table}}
\setcounter{table}{0}
\renewcommand{\theequation}{S\arabic{equation}}
\setcounter{equation}{0}

\newpage

\begin{center}
\begin{Huge}
Supplementary Appendix
\end{Huge}
\end{center}

\vspace{1 cm}

\tableofcontents

\section{Derivation of analytical results presented in the main text}  
\label{sec:SI-analres}
We describe population composition using a stochastic model (see ``Models and methods'' in the main text). This implies that the number $N$ of bacteria in a deme is constantly fluctuating in time. However, because $K\gg 1$ and $g\ll f_S=1$, these fluctuations around the steady-state deme size $N^*$ remain small. At steady state, the division rate balances the death rate, leading to: $f_S(1-N^*/K) =g$ in a population that only comprises S bacteria. This yields
\begin{equation}
   N^* = K\left( 1- g/f_S \right)
\end{equation}
In all the analytical calculations that follow, we approximate deme size $N$ by $N^*$ (see Ref.~\cite{marrec_resist_2020}).

\subsection{Probability of presence of mutants from a lineage destined to go extinct}
\label{subs:ppres}
Let us consider a deme where R mutants have not fixed, in the absence of drug. 
Let us determine the probability $p_\textrm{pres}$ of presence of mutants that arise through mutations upon division of sensitive individuals, and whose lineage is destined to go extinct in the absence of drug, but does not undergo rapid stochastic extinction after drug is added. We follow the method exposed in Ref. \cite{marrec_resist_2020}, which is based on the transition rate matrix of the process. 

\paragraph{Presence of $i$ mutants from a lineage destined to go extinct without drug.} To incorporate the stochastic extinction effect, we need to distinguish each case where exactly $i$ mutants are present, as the probability of stochastic extinction will depend on $i$. The probability of having exactly $i$ mutants present in a well-mixed population with $N^*$ individuals reads $p_R(i)= \tau_R^d(i) \, N^*\mu g$, where $\tau_R^d(i)$ is the average time spent in a state with $i$ mutants by a resistant lineage destined for extinction (known as the sojourn time~\cite{ewens_mathematical_2004}), and $N^*\mu g$ the total mutation rate, with $g$ the death rate and $\mu$ the mutation probability upon division. We have $\tau_{R}^d (i)= -\pi_i/\pi_1 (\mathbf{\tilde{R}}^{-1})_{i1}$, where $\pi_i$ is the probability that resistant mutants go extinct, starting from $i$ of them, while $\mathbf{\tilde{R}}$ is the reduced transition rate matrix for mutants (i.e., the transition rate matrix where rows and columns corresponding to absorbing states are eliminated), see Ref.~\cite{marrec_resist_2020}. This yields the probability that $i$ mutants from a lineage destined to go extinct are present in the deme:
\begin{equation}
p_R(i)= - N^* g \mu \, \frac{\pi_i}{\pi_1} ( \tilde{\mathbf{R}}^{-1} )_{i1}\, .
    \label{eq:ppres0}
\end{equation}

As discussed in the main text, when drug is added to the system, the population can survive if mutants are present when the drug is added. For survival, it is also necessary that the mutant lineages do not quickly go stochastically extinct after drug is added. Let us thus calculate the probability that such a stochastic extinction occurs. 

\paragraph{Probability of stochastic extinction.}
\label{sec:corr-stochext}

Let us consider one resistant bacterium with replication rate $f_R$ and death rate $g$. Let us denote by $e$ the extinction probability of its lineage. This resistant individual can die without dividing with probability $p_D = g/(f_R+g)$, giving an extinction probability of $e_D=1$. It can also divide before dying with probability $p_R = f_R /(f_R+g)$, leading to the extinction probability $e_R = e^2$, assuming independence of lineages, in the branching process framework \cite{harris_branching_1963}. We can then write an equation for the extinction probability $e$ of the lineage of the single resistant bacterium:
    \begin{equation}
        e= p_D \, e_D + p_R \, e_R = \frac{g}{f_R+g} + e^2 \frac{f_R}{f_R+g}.
    \end{equation}
The solutions to this equation are $e=1$ or $e=g/f_R$ if $g<f_R$. Thus, assuming again that the fate of each lineage is independent, we obtain an extinction probability $(g/f_R)^i$ starting from $i$ mutants.

\paragraph{Presence of a mutant lineage destined to go extinct without drug, but that does not undergo stochastic extinction when drug is added.} 
We finally combine Eq~\ref{eq:ppres0} with the stochastic extinction probability to find: 
\begin{equation}
 p_\textrm{pres} = \sum_{i=1}^{N^*} \left[1-(g/f_R)^i\right] p_R(i) = -\sum_{i=1}^{N^*} \left[1-(g/f_R)^i\right] \, N^* g \mu \, \frac{\pi_i}{\pi_1} ( \tilde{\mathbf{R}}^{-1} )_{i1}\, .
	\label{eq:ppres}
\end{equation}
Eq~\ref{eq:ppres} is a good approximation of the survival probability of the population on timescales smaller than the average time of appearance of a successful R mutant, since R mutants that appear before are doomed to go extinct in the absence of drug.

\paragraph{Lifetime of a mutant lineage destined to go extinct without drug.} In the absence of drug, the total average lifetime $\tau_R^d$ of a mutant lineage destined to go extinct can be obtained by summing the sojourn times $\tau_R^d(i)$ expressed above (see also Ref.~\cite{ewens_mathematical_2004}):
\begin{equation}
    \tau_R^d=\sum_{i=1}^{N^*}\tau_R^d(i)=-\sum_{i=1}^{N^*} \frac{\pi_i}{\pi_1} (\mathbf{\tilde{R}}^{-1})_{i1}\,.
\end{equation}
For neutral R mutants, this gives
\begin{equation}
    \tau_R^d=\frac{1}{g}\left(\frac{N^*}{N^*-1}\sum_{i=1}^{N^*-1}\frac{1}{i}-1\right)\approx \frac{1}{g}\log(N^*)\,,
\end{equation}
where we assumed $N^*\gg 1$ to obtain the last expression. This lifetime is only weakly impacted by population size.

\subsection{Appearance of a mutant that fixes in a well-mixed population}
\label{subs:SI-appwm}
In our model, mutations from sensitive to resistant individuals happen at birth with probability $\mu$. Since at equilibrium $f_S\left(1-N^*/K\right) = g$, the average time it takes for a mutant to appear in a population of $N^*$ wild-types is in the rare mutation regime:
\begin{equation}
    \langle t_\textrm{app}(N^*)\rangle = \frac{1}{N^* \mu g}.
    \label{eq:tapp}
\end{equation}
Thus, the average time $\langle t_{af\, W}(N^*) \rangle$ of appearance of a mutant that fixes in a well-mixed population of fixed size $N^*$ is given by:
\begin{equation}
    \langle t_{af\, W}(N^*) \rangle = \langle t_\textrm{app}(N^*)\rangle \times \frac{1}{p_\textrm{fix}(N^*)}= \frac{1}{N^* \mu g\,p_\textrm{fix}(N^*)}.
    \label{eq:taf-wm}
\end{equation}

\paragraph{Cost-free mutants.} The fixation probability of neutral mutants in a population of size $N^*$ is:
\begin{equation}
    p_\textrm{fix}(N^*) = \frac{1}{N^*}\,.\label{eq:pfix_nt}
\end{equation}
Thus, the average time $\langle t_{af\, W} \rangle$ of appearance of a neutral mutant that fixes in a well-mixed population of fixed size $N^*$ is given by:
\begin{equation}
    \langle t_{af\, W} (N^*)\rangle = \langle t_\textrm{app}(N^*)\rangle \times \frac{1}{p_\textrm{fix}(N^*)} = \frac{1}{\mu g}\,.
    \label{eq:taf-wm-nt}
\end{equation}
This implies that for cost-free mutants, the average time of appearance of a successful mutant is independent of the population size: it remains the same for a single deme and a well-mixed population.

\paragraph{Mutants with cost $\delta$.} More generally, the fixation probability of a mutant with fitness cost $\delta$ in a well-mixed population of fixed size $N^*$ described by the Moran process is~\cite{ewens_mathematical_2004}
\begin{equation}
    p_\textrm{fix}(N^*)=\frac{(1-\delta)^{-1}-1}{(1-\delta)^{-N^*}-1}\,.
    \label{eq:general}
\end{equation}
If $\delta\ll 1/N^*$, then to leading order
\begin{equation}
    p_\textrm{fix}(N^*)=\frac{1}{N^*}\,. \label{eq:simple1}
\end{equation}
The mutant is then said to be effectively neutral. Conversely, if $\delta\gg 1/N^*$, then to leading order
\begin{equation}
    p_\textrm{fix}(N^*)=\delta e^{-N^* \delta}\,. \label{eq:simple2}
\end{equation}
Mutants with such a substantial fitness cost have a fixation probability that is exponentially suppressed. 

In Fig \ref{fig:surva-prob}B, as $K\delta=1$, we employ the most general formula for $p_\textrm{fix}$, see Eq~\ref{eq:general}. The cost regimes leading to the simplified expressions of $p_\textrm{fix}$ in Eqs~\ref{eq:simple1} and \ref{eq:simple2} are further discussed in Section \ref{subs:rescost}. 

\subsection{Appearance of the slowest mutant that fixes in a structured population}
\label{subs:slowestfix}
Let us now focus on the appearance of a mutant that fixes in the slowest deme in a population composed of $D$ independent demes which each have steady-state size $N^*$, in the absence of migrations. By \emph{slowest deme}, we mean the deme in which a locally successful mutant (i.e.\ a mutant destined to fix locally in its deme of origin) takes the longest time to appear.

For a mutant to fix in the slowest deme, mutants must have fixed in all others. Let us denote by $t_{af\, S}$ (resp.\ $t_{af\, d}$) the time of appearance of a mutant destined to fix in the slowest deme (resp.\ in any deme). The probability $P(t_{af \, S} \le t)$ that $t_{af\, S}$ is smaller or equal than $t$ reads for any $t$:
\begin{equation}
    P(t_{af\, S} \le t) = [P(t_{af\, d} \le t)]^D.
\end{equation}
The appearance of a locally successful mutant is a Poisson process with rate $\lambda=N^*\mu g\,p_\textrm{fix}(N^*)$ (corresponding to the inverse of Eq~\ref{eq:taf-wm}). Thus, the time $t_{af\, d}$ is exponentially distributed with rate $\lambda$ and probability density
\begin{equation}
  p_d(t_{af\, d}=t)= \lambda e^{- \lambda  t}\,,  
  \label{tafd_exp}
\end{equation}
and we have $P(t_{af\, d} \le t) = 1 - e^{-\lambda t}$. This allows us to express the probability density $p_S(t)$ of appearance of a mutant in the slowest deme:
\begin{equation}
\begin{split}
   p_S(t)dt&=dp( t_{af\, S} \in [t, t+dt])= P(t_{af\, S} \le t+dt)-P(t_{af\, S} \le t) \\
   &=\frac{dP(t_{af\, S} \le t)}{dt}dt = D\left[ P(t_{af\, d} \le t) \right]^{D-1} \, \frac{dP(t_{af\, d}\le t)}{dt} dt \\
    & = D \left[ 1 - e^{-  \lambda  t} \right]^{D-1}  \lambda  e^{- \lambda  t} dt\,.
    \label{eq:pslowest}
\end{split}
\end{equation}
Note that $p_S(t)$ is positive for all $t$ and normalized, as expected.

The average appearance time of a locally successful mutant in the slowest deme is thus given by:
\begin{equation}
   \langle t_{af\, S} (N^*, D)  \rangle = \int_0^{\infty}  t\,p_S(t)\, dt  = \frac{1}{\lambda } \sum_{i=1}^D \frac{1}{i}\,,
   \label{eq:slowest}
\end{equation}
which involves the harmonic number $\sum_{i=1}^D 1/i$.

\subsection{Appearance of the fastest mutant that fixes in a structured population}
\label{subs:fastest}
Let us now consider the \emph{fastest deme}, i.e.\ the deme where a locally successful mutant first appears. By definition, if a successful mutant takes longer than $t$ to appear in the fastest deme, it entails that successful mutants will take longer to appear in all demes. Denoting by $t_{af\,F}$ the time of appearance of a successful mutant in the fastest deme, we have: 
\begin{equation}
    P(t_{af\, F} > t) = [P(t_{af\, d} >t)]^D.
\end{equation}
This allows us to express the probability density $p_F$ of appearance of a successful mutant in the fastest deme:
\begin{equation}
\begin{split}
    p_F(t)dt&=dp(t_{af\,F} \in [t,t+dt]) = - \frac{dP(t_{af\,F} >t)}{dt}dt\\
    &=-D [P(t_{af \, d} > t)]^{D-1} \frac{d P(t_{af\, d} > t)}{dt} dt\\
    & = -D [1 - P(t_{af\, d} \le t)] ^{D-1} \left[ -\frac{dP(t_{af \, d} \le t)}{dt} \right] dt\\
    & = D \lambda e^{-\lambda Dt } dt\,,
\end{split}
\label{eq:papp-fastest}
\end{equation}
with $\lambda=N^*\mu g\,p_\textrm{fix}(N^*)$, as previously defined.
Note that $p_F(t)$ is positive for all $t$ and normalized, as expected.

The average appearance time of a successful mutant in the fastest deme is thus given by (see also Ref.~\cite{bitbol_quantifying_2014}):
\begin{equation}
    \langle t_{af \, F}(N^*, D) \rangle = \int_0^{\infty} t\, p_F(t)\, dt  =  \frac{1}{D\lambda }\,.
\end{equation} 
In the case of cost-free mutants, using Eq~\ref{eq:pfix_nt} gives
\begin{equation}
    \langle t_{af \, F}(N^*, D) \rangle =  \frac{1}{D\mu g}\,,
\end{equation}
In the case where resistance carries a cost, using Eq~\ref{eq:general} gives
\begin{equation}
    \langle t_{af\, F}(N^*, D) \rangle = \frac{1}{D N^* \mu g} \frac{(1-\delta)^{-N^*}-1}{(1-\delta)^{-1}-1}\,.
    \label{eq:tapp-fastest-cost}
\end{equation}

\subsection{Survival probability of the population with $\gamma=0$}
\label{subs:surv-ext}
When there is no migration between demes, i.e.\ when $\gamma=0$, the survival probability $p_{s\, \gamma=0}(D)$ of a population comprising $D$ demes, is related to the survival probability $p_{s\, d}$ of a deme through:
\begin{equation}
    p_{s\, \gamma=0} (D)   = 1- (1-p_{s\,d})^D.
    \label{eq:combinations-ddemes}
\end{equation}
Indeed, denoting by $p_{e\, \gamma =0 }$ the probability of extinction of all demes, we have:
\begin{equation}
     p_{s\, \gamma = 0} (D)=1-p_{e\, \gamma=0}(D).
\end{equation}
Let $p_{e\, d}$ be the probability that the population in one deme becomes extinct. When $\gamma =0$, extinctions in different demes are independent events with the same probability, thus:
\begin{equation}
    p_{s\, \gamma = 0} (D)= 1- p_{e\,\gamma =0}(D) =1- \left( p_{e\, d} \right)^D = 1- (1-p_{s\,d})^D\,.
    \label{eq:psurv-ddemes}
\end{equation}

We checked that Eq~\ref{eq:combinations-ddemes} was satisfied in our numerical simulations, within the errors given by the standard error of the proportion.

\subsection{Closed form of the survival probability for well-mixed and fully subdivided populations}
\label{sec:purv_analytical}
In our structured populations, mutants appear at random in any deme. The population can survive in two distinct ways: \begin{itemize}
\item Due to the presence of successful lineages that have fixed locally or globally in the population, and allow survival when drug is added. 
\item Due to the presence of a mutant lineage destined for extinction in the absence of drug at the time $t_{\textrm{add}}$ when drug is added. This mutant lineage can proliferate in the presence of drug. The population then survives, unless this mutant lineage undergoes a quick stochastic extinction.
\end{itemize}
The survival probability can be then written by summing over these two distinct scenarios:
\begin{equation}
\begin{split}
 p_{\textrm{surv}}(t_{\textrm{add}}) &= p_\textrm{succ, surv}(t_{\textrm{add}}) + p_{\textrm{no\, succ, surv}} (t_{\textrm{add}}) \\
 &=p_{\textrm{succ}}(t_{\textrm{add}})\, p_{\textrm{surv} | \textrm{succ}} + \left[1-p_{\textrm{succ}}(t_{\textrm{add}})\right]  p_{\textrm{surv} |\textrm{no\, succ}} \\
 &= p_{\textrm{succ}}(t_{\textrm{add}}) + \left[1-p_{\textrm{succ}}(t_{\textrm{add}}) \right] p_{\textrm{pres}}\,.
\end{split} 
\label{eq:psurv-analy-SI}
\end{equation}
In the first line, we denoted by $p_{\textrm{succ, surv}}(t_{\textrm{add}})$ the probability that a successful mutant has appeared by $t_{\textrm{add}}$ and leads to population survival, and by $p_{\textrm{no\, succ, surv}} (t_{\textrm{add}})$ the probability that no successful mutant has appeared  at $t_{\textrm{add}}$ but that the population survives (thanks to the presence of a mutant lineage that was destined for extinction). In the second line, $p_{\textrm{succ}}(t_{\textrm{add}})$ is the probability that a successful mutant has appeared by $t_{\textrm{add}}$. The probability $ p_{\textrm{surv} | \textrm{succ}}$ of survival conditioned on the presence of a successful mutant is one, while the probability $ p_{\textrm{surv} |\textrm{no\, succ}}$ of survival conditioned on the absence of any successful mutant is $p_{\textrm{pres}}$, given by Eq~\ref{eq:ppres}.

Let us now calculate $p_{\textrm{succ}}(t_{\textrm{add}})$ in a well-mixed population of fixed size $N^*$. Given that the appearance of a successful mutant is a Poisson process with rate $\lambda =N^* \mu g p_{\textrm{fix}}(N^*)$, the associated time of appearance is distributed according to Eq \ref{tafd_exp}. Thus, $p_{\textrm{succ}}(t_{\textrm{add}})$ reads:
\begin{equation}
  p_{\textrm{succ}}(t_{\textrm{add}}) = \int_0^{t_{\textrm{add}}} \lambda e^{-\lambda t} dt = 1 - e^{-\lambda t_{\textrm{add}}}.
  \label{eq:psucc-general}
\end{equation}
For neutral mutants in a well-mixed population of size $N^*$, this reads:
\begin{equation}
     p_\textrm{succ}(t_\textrm{add})=1-e^{-\mu g t_\textrm{add}},
    \label{eq:psucc-W-nocost}
\end{equation}
while for deleterious mutants in a well-mixed population of size $N^*$ we have:
\begin{equation}
    p_\textrm{succ}(t_\textrm{add}) =1 - \exp\left(-N^*\mu g \frac{(1-\delta)^{-1}-1}{(1-\delta)^{-N*}-1} t_\textrm{add}\right).
    \label{eq:psucc-W-cost}
\end{equation}

Let us now consider a structured population, and ask whether a locally successful mutant has appeared by $t_\textrm{add}$. For the appearance time of successful mutants in the fastest deme of a structured population, the probability density is given by Eq~\ref{eq:papp-fastest}, yielding:
\begin{equation}
     p_\textrm{succ}(t_\textrm{add}) = \int_0^{t_\textrm{add}} D \lambda e^{-\lambda D t} dt = 1 - e^{D \lambda t_\textrm{add}}.
\end{equation}
If mutants are neutral, we have:
\begin{equation}
    p_\textrm{succ}(t_\textrm{add}) = 1-e^{-D\mu g t_\textrm{add}},
    \label{eq:psucc-S-nocost}
\end{equation}
and if they carry a resistance cost:
\begin{equation}
     p_\textrm{succ}(t_\textrm{add}) = 1 - \exp\left(-D N^*\mu g \frac{(1-\delta)^{-1}-1}{(1-\delta)^{-N*}-1} t_\textrm{add}\right) .
    \label{eq:pucc-S-cost}
\end{equation}

\section{Description of the simulations performed in the main text}

\subsection{General simulation approach}
\label{subs:SI-ssetup}
Our simulations are based on the Gillespie algorithm \cite{gillespie_exact_1977, gillespie_general_1976, higham_modeling_2008}. Let us consider a clique population structure composed of $D$ demes, labeled with $i\in [1, D]$. Let $S_i$ (resp.\ $R_i$) denote one sensitive (resp.\ resistant) individual in deme $i$. The system obeys the following reaction network:
\begin{equation}
\begin{split}
&S_i  \xrightarrow[ ]{f_S\left(1-\frac{N}{K}\right)(1-\mu)} 2 S_i\,,\\
&S_i  \xrightarrow[ ]{f_S\left(1-\frac{N}{K}\right)\mu} S_i+R_i\,,\\
&S_i \xrightarrow[ ]{g} \emptyset\,,\\
&S_i \xrightarrow[]{\gamma} S_j\,\,\, \textrm{for all}\,\,\, j\neq i\,,\\
&R_i  \xrightarrow[ ]{f_R\left(1-\frac{N}{K}\right)} 2R_i\,,\\
&R_i  \xrightarrow[ ]{g} \emptyset\,,\\
&R_i \xrightarrow[]{\gamma} R_j\,\,\, \textrm{for all}\,\,\, j\neq i\,.
\end{split}
\end{equation}

The action of the biostatic drug is modeled through a change of the fitness of sensitive individuals, $f_S$, as follows:
\begin{equation}
    f_S = \begin{cases}
        1 \hspace{0.3 cm} \text{in the absence of drug,}\\
        0 \hspace{0.3 cm} \text{after biostatic drug addition.}
    \end{cases}
\end{equation}

\subsection{Simulation runs and analyses}
In general, we run simulations as follows. For each drug addition time $t_\textrm{add}$ considered, we run multiple simulation replicates. In each of them, we initialize the system with $0.1\times K$ sensitive individuals in each deme (or $0.1\times DK$ in the well-mixed population), let the system evolve until $t_\textrm{add}$ and save its state. Then we add the biostatic drug, and we let the system evolve until either extinction of the population or full system colonization by resistant individuals. If resistance fixes before drug addition, we save the state at this point and stop the simulation. When stopping the simulation, in all cases described above, we save the stopping time. Because of stochastic fluctuations around the mean, we consider that mutants have colonized one deme (resp. the whole population in the well-mixed case) if the population size is greater than $0.9\times K$ (resp. $0.9\times DK$).

Below, we describe in more detail how simulations are performed and analyzed for each of the main figures.

\paragraph{Survival probability.} In Fig \ref{fig:surva-prob}, we calculate the survival probability as the fraction of simulation replicates where bacteria survive the application of the drug, in the sense that one deme comprises at least $0.9\times K$ mutants. Note that this allows us to exclude cases in which mutants quickly go stochastically extinct after drug addition. In the structured populations, survival is intended to be in at least one of the demes. 

\paragraph{Average and variance of mutant numbers at the population level.} In Fig~\ref{fig:composition-T5e5}A-B, for each time and each structure, we perform simulation replicates in the absence of drug. For each of these replicates, we count the number of mutants in the population. We then calculate the average and variance of the total number of mutants across the different replicates. 

\paragraph{Average and variance of mutant numbers at the deme level.} 
In Fig \ref{fig:ave-var-deme}, for each time and each structure, we perform simulation replicates in the absence of drug. For each replicate we choose one deme (deme 1 in practice in Fig \ref{fig:ave-var-deme}, but the result is similar for other demes given the symmetry of the clique). We calculate the average and variance of the number of mutants in that deme across stochastic realizations.

\paragraph{Population composition without drug.} In Fig~\ref{fig:composition-T5e5}C, we analyze the composition of the population before the addition of drug. In the state of the system saved before environmental change, we focus on the number of mutants, and partition the outcomes of our stochastic simulations into the four categories described in the main text (``Only R'', ``Big R pop.'', ``Small R pop.'', ``No R''). We report the fraction of replicates falling in each of these categories.

\paragraph{Number of demes where resistance has fixed.} In Fig~\ref{fig:fixR}, for each time and a structure, we perform multiple replicates. For each of them, we calculate the number of demes where the mutant type has fixed before drug addition. We report a histogram of these number of demes across replicates.

\paragraph{Time until colonization of next deme.} In Fig~\ref{fig:timescales}A, we perform simulations to assess the time until colonization of next deme. These simulation are set up slightly differently than others, to focus on the colonization process. We initialize one of the demes with $N=0.9\times K$ mutants, all others with $N=0.9\times K$ wild-types, mimicking fixation of mutants in one deme before drug addition. We add the biostatic drug at $t=0$ in this simulation, thus modeling the case where one deme has fixed resistance before drug addition. We then let the system evolve until overall colonization of the structured population by resistant mutants. For each $k\geq 1$, we save the times of successful colonization of the $(k+1)^{th}$ deme, given that $k$ demes were already colonized by mutants. We define successful colonization of a deme as reaching a number of mutants $0.9\times K$. We then obtain the values of $\langle t_\textrm{c mig} (k) \rangle$ as the differences between the successive times we recorded. 

\paragraph{Time until overall colonization.} In Fig~\ref{fig:timescales}B, we report the time until overall colonization of the structured population by resistant mutants. When saving the state of the system after overall mutant colonization (resp.\ extinction), we also record the time it took for overall colonization (resp.\ extinction). Thus, we can calculate the mean of the overall colonization time for each $t_\textrm{add}$ and each structure. 
Note that the fixation of resistance may happen before drug addition (this happens in particular for addition times longer than $\langle t_{af \, S} \rangle $). In this case, we consider that colonization happens when resistant individuals reach fixation in all demes and there are at least $0.9\times K$ resistant individuals.

\section{Clique population structure with sensitive inoculum} 
\label{sec:pop-analysis}

\subsection{Clique population structure with different numbers of demes}
\label{sec:moredemes}
\begin{figure}[htb!]
    \centering
    \includegraphics[width=0.6\linewidth]{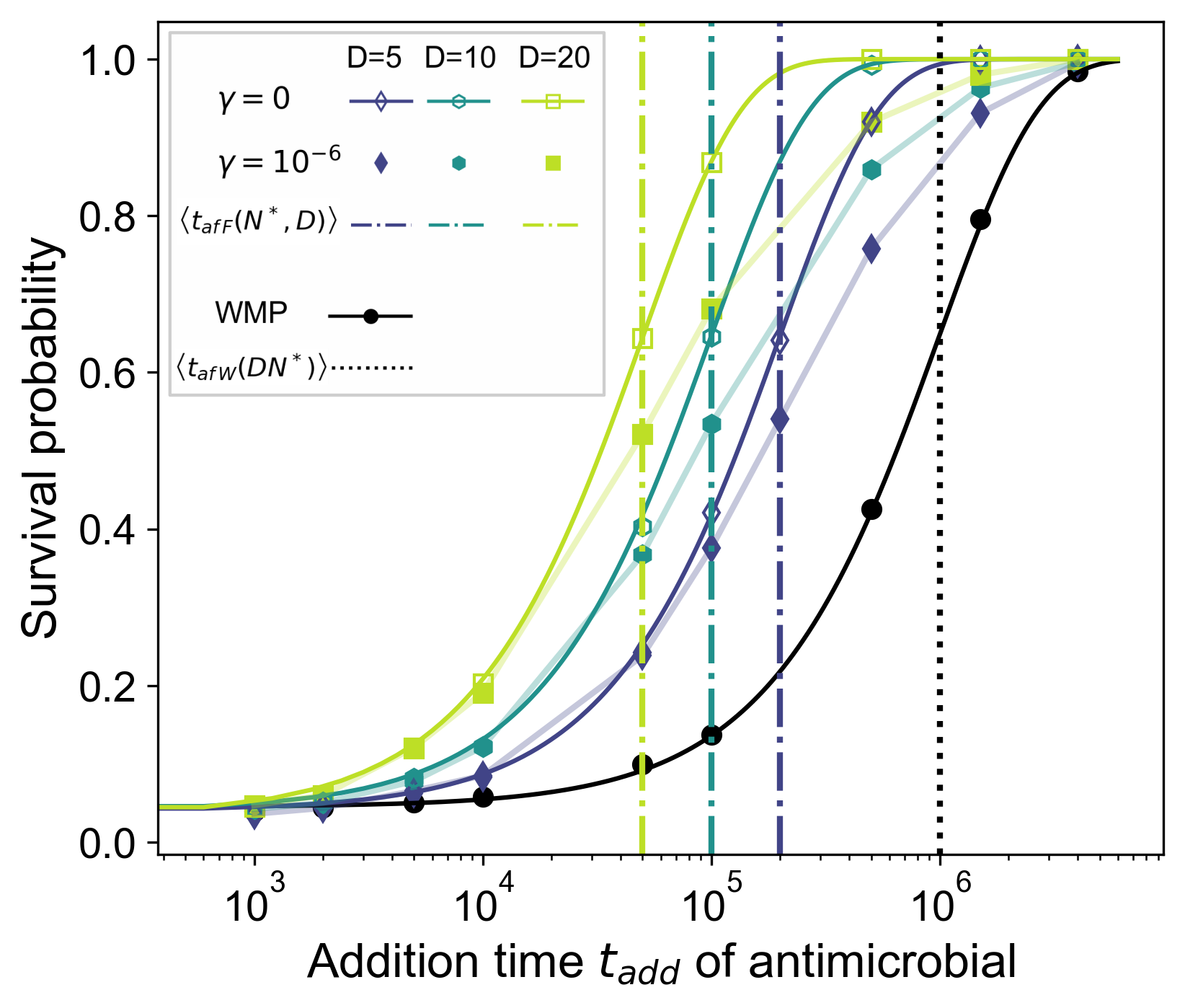}
    \caption{\textbf{Survival probability of a bacterial population with neutral mutants for varying degree of subdivision at a fixed total population size.} Survival probability as a function of the treatment addition time $t_\textrm{add}$. Three levels of subdivision of a population with total carrying capacity $DK=1000$, into 5, 10, and 20 demes, are shown for two values of $\gamma$. Results for the well-mixed population (``WMP") are shown for reference. In all cases, we determined the survival probability from the fraction of $10^4$ simulations where the population survived after drug application. The vertical dash-dotted lines denote the average time of appearance of a locally successful mutant in the fastest deme of each structure (Eq~\ref{eq:fastest}). The vertical dotted line represents the average appearance time of a successful mutant in a well-mixed population (Eq~\ref{eq:tfixgen}). Solid lines for the fully subdivided ($\gamma=0$) and well-mixed populations are analytical predictions from Eq~\ref{eq:psurv-analytical}. Thin lines connecting data points are guides for the eye. Parameter values: $f_S=1$ without drug, $f_S=0$ with drug, $f_R=1$ (no cost of resistance), $g = 0.1$, $\mu=10^{-5}$.}
    \label{fig:moredemes}
\end{figure}

To assess the impact of the degree of subdivision of a population on rescue by resistance, we extend the study of the survival probability with neutral R mutants in Fig \ref{fig:surva-prob}A to different numbers of demes. 
Fig \ref{fig:moredemes} shows the impact of varying the degree of subdivision of a population of fixed total size $DK$, but composed of a different number of demes ($D=5, 10, 20$). We observe that higher subdivision yields higher probabilities of survival.

\newpage

\subsection{Impact of a cost or of a benefit of resistance}
\label{subs:rescost}

In Fig~\ref{fig:surva-prob}, we considered the case where the resistant mutant is neutral in the absence of drug and the case where it carries a moderate cost of resistance $\delta=1/K$. Here, we discuss different regimes of cost of resistance, and present simulation results for a higher cost of resistance. 
Besides, if antibiotic is already present at relatively low doses in the environment before the addition of drug, it can induce a benefit of resistance before drug is added. We briefly discuss this case at the end of this section.

\paragraph{Different regimes of cost.}
Let us consider different regimes of cost for R mutants. 

As discussed in the main text, for neutral resistant mutants ($\delta=0$), $p_\textrm{fix}=1/N^*$, so Eq~\ref{eq:tfixgen} gives $\langle t_{af\, W} (N^*)\rangle =1/(\mu g)$. Importantly, this result does not depend on population size $N^*$. Therefore, it holds both for each deme in a structured population and for a well-mixed population of steady-state size $DN^*$: $\langle t_{af\, W} (DN^*)\rangle =1/(\mu g)$. 
Meanwhile, Eq~\ref{eq:fastest} gives $\langle t_{af\, F} (N^*,D)\rangle =1/(D\mu g)=\langle t_{af\, W} (N^*)\rangle/D$. Therefore, a locally successful mutant appears in a structured population $D$ times faster than a successful mutant in a well-mixed population with same total steady-state size $DN^*$.

For mutants with a substantial cost of resistance $\delta\gg 1/K$, $p_\textrm{fix}=\delta e^{-N^*\delta}$ (see Eq~\ref{eq:simple2}), and thus $\langle t_{af\, W} (N^*)\rangle = e^{N^* \delta}/(N^*\mu g \delta)$, which is exponentially longer than in the neutral case and satisfies $\langle t_{af\, W} (N^*)\rangle\ll \langle t_{af\, W} (DN^*)\rangle$. Therefore, $\langle t_{af\, F} (N^*,D)\rangle=\langle t_{af\, W} (N^*)\rangle/D\ll \langle t_{af\, W} (DN^*)\rangle$. 

Finally, mutants with intermediate cost satisfying $1/(DK)\ll\delta\ll1/K$ are effectively neutral in demes of size $N^*$, leading to $\langle t_{af\, F} (N^*,D)\rangle =1/(D\mu g)$ as in the neutral case (see Eq~\ref{eq:simple1}). However, they are substantially deleterious in a well-mixed population of size $DN^*$, leading to $\langle t_{af\, W} (DN^*)\rangle = e^{DN^* \delta}/(DN^*\mu g \delta)$, which is exponentially larger than $\langle t_{af\, F} (N^*,D)\rangle$.

\paragraph{Strong cost of resistance.} 

\begin{figure}[htbp]
    \centering
    \includegraphics[width=0.6\linewidth]{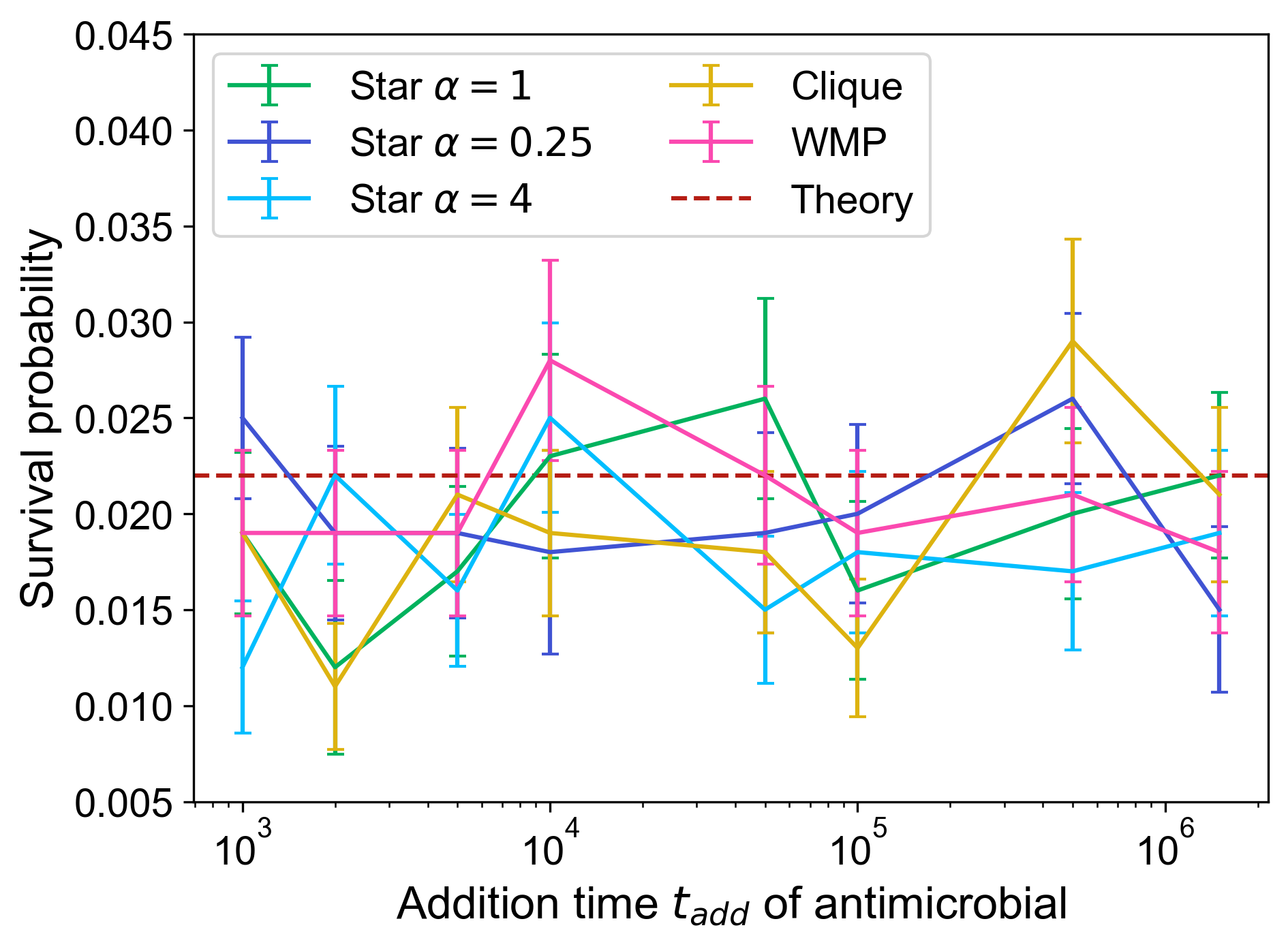}
    \caption{\textbf{Survival probability of a structured bacterial population with a cost of resistance.} The survival probability of the population to the addition of biostatic drug is shown versus the drug addition time. We consider a clique and a star with different values of migration asymmetry $\alpha$. The case of the well-mixed population with the same total size (``WMP") is shown for reference. The migration rate is $\gamma=10^{-6}$ in the clique, while in the star we follow the convention in Section \ref{sec:diff-popstruct} to compare with a clique of a given migration rate $\gamma$, leading to $\gamma_O=(D-1)\gamma$ and $\gamma_I = \alpha \gamma_O$. Simulation results are obtained from $10^3$ replicates, with error bars representing 95\% confidence intervals. Red horizontal dashed line (``Theory"): analytical prediction from Eq~\ref{eq:ppres}. Parameter values: $K=200$, $D=5$, $f_S=1$ without drug, $f_S=0$ with drug, $f_R=0.9$ with and without drug, $g = 0.1$, $\mu=10^{-5}$.}
    \label{fig:psurv-09}
\end{figure}

With a strong cost of resistance $\delta$ satisfying $\delta\ll 1$ and $N^*\delta\gg 1$ (satisfied in Fig~\ref{fig:psurv-09}), the probability of fixation of one mutant in a deme of size $N^*$ is approximately $\delta e^{-N^*\delta}$ (cf. Eq~\ref{eq:simple2}). It is thus exponentially suppressed, and the time to appearance of a successful mutant in a well-mixed population of size $N^*$ is exponentially longer than in the neutral case. With the parameter values of Fig~\ref{fig:psurv-09}, the appearance time of a successful mutant in a well-mixed population of size $DN^*$ is $\langle t_{af\, W} (DN^*) \rangle \approx 1.3 \times 10^{43}$, while the appearance time of the fastest locally successful mutant in a population of $D$ demes of size $N^*$ each is $\langle t_{af\, F} (N^*, D) \rangle \approx 1.3 \times 10^8$. Therefore, fixation events can be neglected in the range of $t_\textrm{add}$ considered previously for neutral mutants and mutants with moderate cost (see Fig \ref{fig:surva-prob} for the clique, and in Fig \ref{fig:latt-star-cl} for the star). 
Thus, in this range of $t_\textrm{add}$, we can estimate the survival probability from the probability of presence of mutant lineages destined for extinction, introduced in Eq~\ref{eq:ppres}. This analytical prediction agrees with our simulation results, see Fig \ref{fig:psurv-09}. We do not observe a statistically significant impact of population structure on the survival probability of the population, in this range of $t_\textrm{add}$. Recall however that if times comparable to the mean fixation time of a successful mutant in a deme (or larger) were considered, we predict that spatial structure would favor population survival for strong costs of resistance. We do not consider this further because of the extremely long times involved.

\paragraph{Benefit of resistance.} Let us now consider the case where the resistant mutant has a fitness $f_R=1+s$ with a selective advantage $s>0$ in the absence of drug. Assuming $s\ll 1$ but $N^* s\gg 1$, the probability of fixation of one mutant in a deme of size $N^*$ is approximately $s$ in the Moran model~\cite{ewens_mathematical_2004}. 
As this is independent of $N^*$, the same holds for a well-mixed population of size $DN^*$. Thus, in this case, the average time to appearance of a successful mutant is the same in a deme and in a well-mixed population. This entails that the effect of spatial structure we evidenced for neutral and deleterious resistant mutants does not extend to beneficial ones.

\subsection{Growth of mutant number in structured and well-mixed populations}
\label{subs:growth}

\begin{figure}[htbp]
    \centering
    \includegraphics[width=0.95\linewidth]{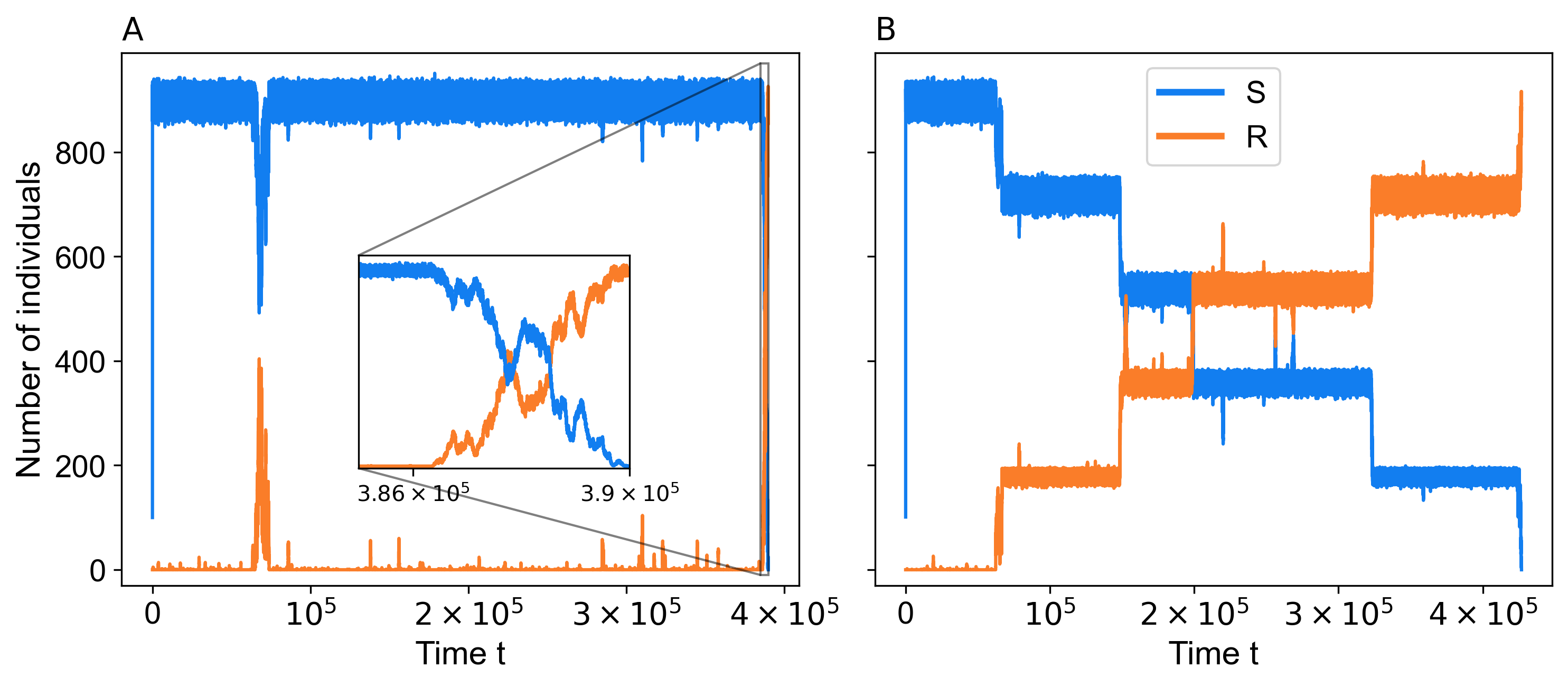}
    \caption{\textbf{Dynamics of population in a well-mixed and a structured population.} We present one trajectory from a single simulation realization in the absence of drug for a well-mixed population in Panel A, and one for a structured population with same total size in Panel B. In both cases, we show the number of sensitive (S) and resistant (R) individuals versus time. Parameter values: $K=200$, $D=5$, $\mu=10^{-5}$, $f_S=f_R=1$ (neutral mutants), $g=0.1$, $\gamma=10^{-7}$.}
    \label{fig:profiles}
\end{figure}

Starting from an inoculum of sensitive bacteria, resistant mutants may appear and fix in the absence of drug. In the case of a structured population with small migration rates, they locally fix deme after deme. 
This results in a substantial difference in the growth pattern of the number of resistant individuals between a well-mixed and a structured population. 
We present one specific trajectory of the number of sensitive and neutral resistant individuals in a well-mixed population in Fig \ref{fig:profiles}A and one in a structured population in Fig \ref{fig:profiles}B. 
In the structured population, a locally successful mutant appears on average earlier than a successful one in the well-mixed population. However, it then takes more time for resistance to fix in the whole population, as this involves either independent appearance of other mutants or migration of resistant mutants from deme to deme. Accordingly, Fig \ref{fig:profiles}A features a rapid but late growth of mutant fraction in the well-mixed population. Meanwhile, Fig \ref{fig:profiles}B, we observe that the growth of mutant fraction in the structured population features $D-1$ intermediate plateaus, as fixation occurs deme after deme. Note that the sequential fixation pattern shown in Fig \ref{fig:profiles} is also observed for a successful mutant that carries a cost of resistance, but the involved timescales differ (see Supplementary Appendix Section \ref{subs:rescost}).

Despite these differences in the individual trajectories of the number of mutants, their average over many replicates is the same whatever the migration rate for neutral mutants, see Fig \ref{fig:composition-T5e5}A. However, these differences in trajectories are at the root of those observed in the variance across replicates shown in Fig \ref{fig:composition-T5e5}B. The well-mixed population features more variance in mutant numbers across replicates than structured populations with small migration rates, because of the variability of the appearance time of a successful mutant. 

\subsection{Stochastic extinction}
\label{subs:stochext}
The probability of stochastic extinction starting from $j_0$ individuals in a population can be computed by formally integrating the Master equation with the initial condition $j=j_0$:
\begin{equation}
    P_0(t) = \left( e^{\mathbf{R}t}\right)_{0j_0},\label{eq:P0R}
\end{equation}
where $\mathbf{R}$ is the transition rate matrix describing the population dynamics, see Ref. \cite{marrec_resist_2020}. In particular, for a population with a carrying capacity $K=10^2$ composed of individuals with fitness $f_R=1$, the probability of rapid initial extinction starting from 10 individuals or more is negligible (lower than $10^{-9}$) \cite{marrec_resist_2020}. (Note that this extinction probability can also be calculated in the branching process approximation of Section \ref{subs:ppres}, giving $(g/f_R)^{10} = 10^{-10}$ here). This is why, in our categorization of results used in Fig \ref{fig:composition-T5e5}, we called ``Large R population'' the case where the number of mutants is at least 10: the possibility of stochastic extinction can be neglected in this case.

\subsection{Dynamics at the single-deme level}
\label{subs:dyndeme}

\begin{figure}[htbp]
        
        \centering\includegraphics[width=0.95\linewidth]{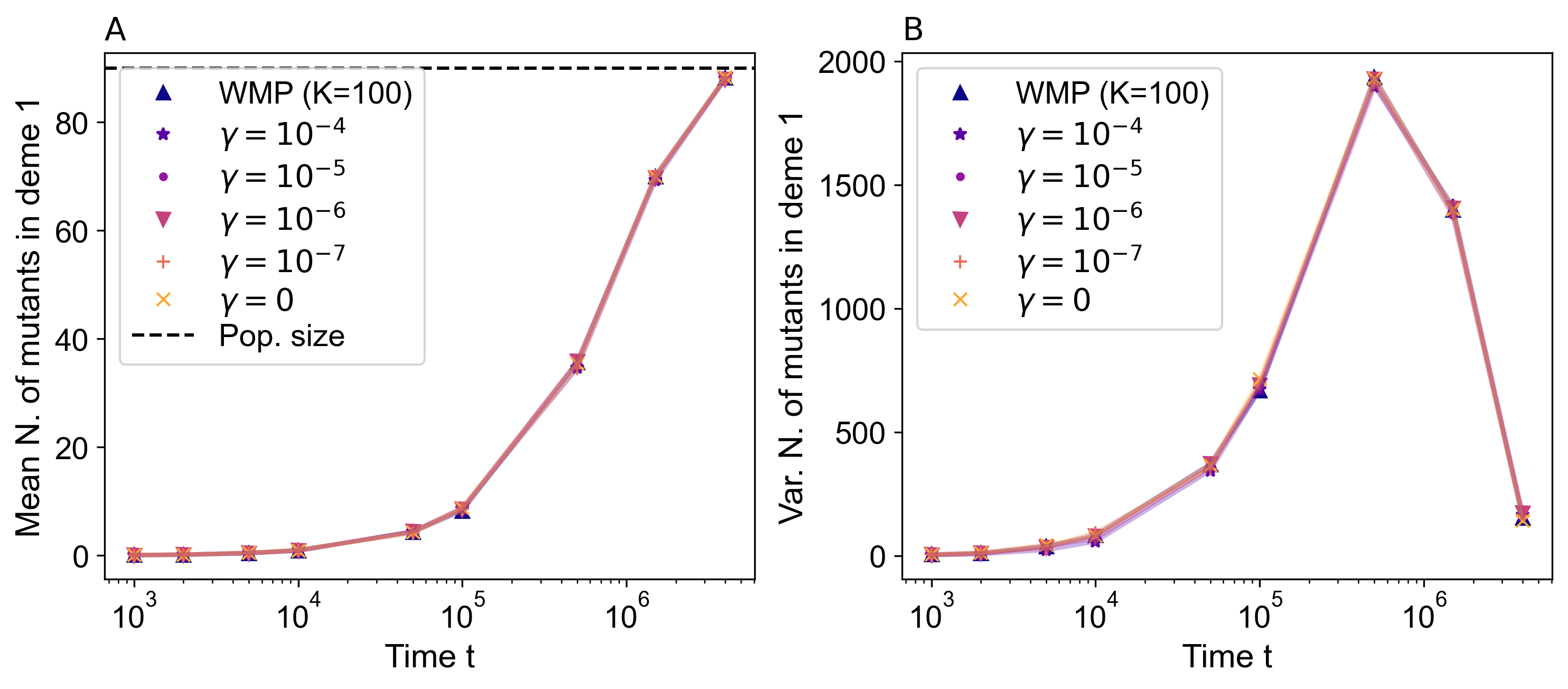}
    
    \caption{\textbf{Mean and variance of the number of mutants at the single-deme level.} Panel A: the mean number of mutants in deme 1 is shown as a function of time in the absence of drug, for different values of the migration rate. Panel B: the variance across simulation replicates of the number of mutants in deme 1 is shown as a function of time in the absence of drug, for different values of the migration rate. Data is obtained from $10^4$ simulation replicates in each case. The case of an isolated deme, i.e.\ of a well mixed population with carrying capacity $K$ (``WMP ($K=100$)"), is shown for reference. Parameter values (in both panels): $K=100$, $D=10$, $f_S=f_R=1$ (neutral mutants),  $g = 0.1$, $\mu=10^{-5}$. }
    \label{fig:ave-var-deme}
\end{figure}

In Fig \ref{fig:composition-T5e5}, we analyzed the time evolution of mean and variance across replicates of the total number of neutral mutants in the population. In Fig \ref{fig:ave-var-deme}, we analyze these quantities at the single-deme level, always for neutral mutants. Since the clique is a symmetric structure such that all demes are equivalent, we select deme 1 in each of our simulation replicates for this analysis. Fig \ref{fig:ave-var-deme}A shows the average number of mutants in this deme as a function of time. This quantity is the same in the structures with different migration rates, as is the case at the population level (see Fig \ref{fig:composition-T5e5}A). Conversely, Fig \ref{fig:ave-var-deme}B shows that the variance of the number of mutants in one deme differs from its population-level counterpart (see Fig \ref{fig:composition-T5e5}B). Indeed, at the deme level, there is no visible difference between different population structures as far as the variance across replicates of the number of mutants in a deme is concerned, see Fig \ref{fig:ave-var-deme}B. Thus, the local neutral mutant dynamics are the same for all migration rates. It is the global dynamics at the population level, combining all demes, which gives rise to the impact of spatial structure that we evidenced in this work. 

\subsection{Population composition versus time for different migration rates}
\label{subs:compodeme}

\begin{figure}[htb!]
    \centering
    \includegraphics[width=\textwidth]{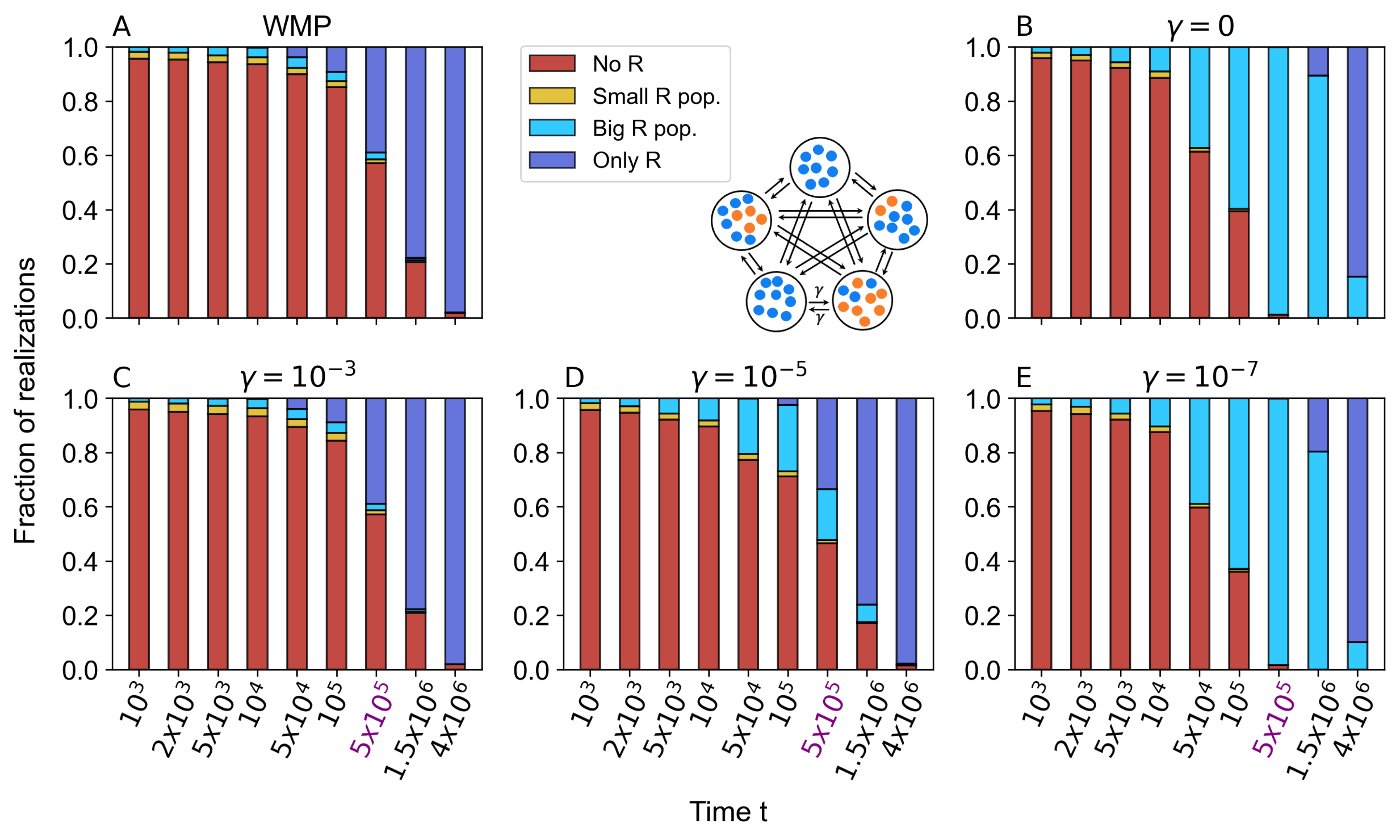}
    \caption{\textbf{Population composition versus time for different population structures.} Population composition in the absence of drug in the well-mixed population (``WMP", Panel A), in the fully subdivided population (``SP", Panel B), and in the clique-structured population with same total size for three values of the migration rate $\gamma$ (Panels C-E). The four categories of population composition reported here are based on the number of mutants per deme. They are the same as in Fig \ref{fig:composition-T5e5}C, and are defined in the caption of that figure. The x-axis tick corresponding to $t=5\times 10^5$, considered in Fig \ref{fig:composition-T5e5}C, is highlighted in purple. Data is obtained from $10^4$ simulation replicates in each case. Parameter values (in all panels): $K=100$, $D=10$, $f_S=f_R=1$, $g = 0.1$, $\mu=10^{-5}$.}
    \label{fig:histo-comp}
\end{figure}

In Fig \ref{fig:composition-T5e5}C, we analyzed the composition of the population at the time featuring the largest inter-replicate variance for the well-mixed population, namely $t=5\times 10^5$, in spatially structured populations with different migration rates, in the case of neutral R mutants. In Fig \ref{fig:histo-comp}, we report how this composition evolves in time, in the absence of drug, for spatially structured populations with different migration rates. As in Fig \ref{fig:composition-T5e5}C, we focus on the number of mutants in each deme to categorize the population composition. 

Recall that here, the population is composed of wild-type sensitive bacteria and of neutral resistant mutants that arise with probability $\mu$ upon division of wild-types. Fig \ref{fig:histo-comp} shows that when the per capita migration rate $\gamma$ is decreased, there is a longer phase where the population has an intermediate composition, with some demes harboring a substantial number of resistant mutants (``Big R pop." category). This is consistent with our above discussion in Section \ref{subs:growth} and with our results in Fig \ref{fig:profiles}. The fixation of mutants occurs gradually, deme by deme, in structured populations with small migration rates. Furthermore, substantial mutant numbers exist earlier in these structured populations. We further observe in Fig \ref{fig:histo-comp} that a per capita migration rate $\gamma=10^{-3}$ results in an evolution of population composition akin to that of a well-mixed population, a per capita migration rate $\gamma=10^{-7}$ yields an evolution of population composition similar to that of a fully subdivided population with $\gamma=0$. Structures with intermediate values of $\gamma$ feature an intermediate evolution of population composition.

\subsection{Colonization timescales after drug is added}
\label{sec:colSI}

In the main text, we discuss the time $\langle t_\textrm{c mig} \rangle$ for R mutants to colonize the next empty deme after drug is added, see Eq \ref{eq:taufmig}. In Table \ref{tab:timescales}, we report values of $\langle t_\textrm{c mig} (k)\rangle$ when $k=1$ or 5 demes are already mutant, for different values of the migration rate $\gamma$, using the same parameter values as in Fig~\ref{fig:timescales}. With these parameters, the decay time upon drug addition of a well-mixed population comprising $N^*$ S bacteria is $\tau_S=50.8$. Recall that the smallest value of $\langle t_\textrm{c mig} (k)\rangle$ is obtained when $k=D/2=5$. Thus, the results of Table \ref{tab:timescales} show that for $\gamma=10^{-6}$, which is the value used in Fig \ref{fig:timescales}, we have $\tau_S\ll \langle t_\textrm{c mig} (k)\rangle$ for all $k$.

\begin{table}[ht]
\centering

\renewcommand{\arraystretch}{1.4} 
\setlength{\tabcolsep}{10pt} 
\begin{tabular}{@{}lcccc@{}}
\toprule
 & \textbf{\boldmath$\gamma=10^{-4}$} & \textbf{\boldmath$\gamma=10^{-5}$} & \textbf{\boldmath$\gamma=10^{-6}$} & \textbf{\boldmath$\gamma=10^{-7}$} \\
\midrule
$\langle t_\textrm{c mig} (k=1) \rangle $ & $13.7$ & $1.37 \times 10^2$ & $1.37 \times 10^3$ & $1.37 \times 10^4$ \\
$\langle t_\textrm{c mig}(k=5) \rangle $ & $4.93$ & $49.3$  & $4.93 \times 10^2$ & $4.93 \times 10^3$ \\
\bottomrule
\end{tabular}
\caption{Numerical evaluation of $\langle t_\textrm{c mig}\rangle$ from Eq~\ref{eq:taufmig} for $k=1$ and $k=5$. Parameter values: $K=100$, $D=10$, $g=0.1$, $f_R=1$ (neutral mutants), as in Fig~\ref{fig:timescales}.}
\label{tab:timescales}
\end{table}

The total colonization time $\langle t_\textrm{c tot}\rangle$ can be obtained by summing $\langle t_\textrm{c mig} \rangle$ over the $D-1$ steps needed to sequentially colonize the structured system. Using Eq \ref{eq:taufmig}, we thus obtain:
\begin{equation}
    \langle t_\textrm{c tot}\rangle = \sum_{k=1}^{D-1} \langle t_\textrm{c mig} (k) \rangle   = \frac{2 (\Gamma + \psi(D))}{D\gamma N^*(1-g/f_R)}\,,
    \label{eq:Tftot}
\end{equation}
where $\Gamma$ is the Euler gamma constant, while $\psi$ is the digamma function.

\section{Lattice, star and line structures with sensitive inoculum} 
\label{sec:diff-popstruct}

Our work mainly focuses on a minimal model of spatially structured populations where all demes are equivalent and connected to one another by identical migration rates. This structure is known as the island model, the clique or the fully connected graph. Let us now extend our study to different graph structures. 

\begin{figure}[htb!]
    \centering
    \includegraphics[width=\textwidth]{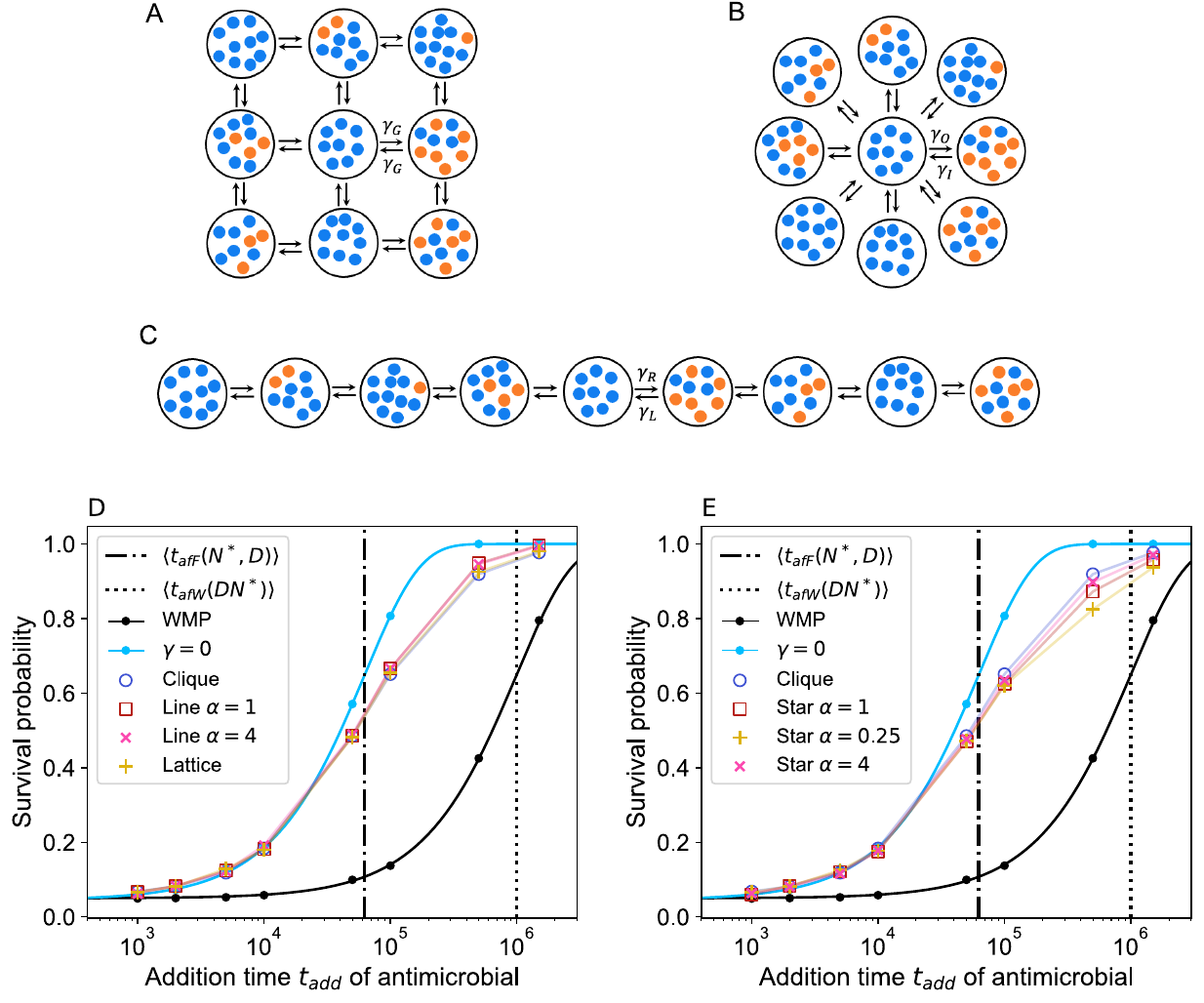}
    \caption{\textbf{Survival probability of a bacterial population to biostatic drug, for different spatial structures.} Panels A, B, C: Schematics of the spatial structures considered, shown with $D=9$ demes for visualization ease: a square lattice with migrations $\gamma_G$ to each nearest neighbor (A), a star with and migration rates $\gamma_O$ from the center to a leaf and $\gamma_I= \alpha \gamma_O$ from a leaf to the center (B), and a line with migration rates $\gamma_L$ to the left and $\gamma_R = \alpha \gamma_L$ to the right (C). Panels D, E: Survival probability of a population starting from a sensitive inoculum versus drug addition time $t_\textrm{add}$, for different spatial structures: line with different migration asymmetries and lattice (D), star with different migration asymmetries (E). Results for a fully subdivided population with no migrations (``$\gamma=0$"), a clique, and a well-mixed population are shown in both panels for reference. For the fully subdivided and well-mixed populations, the prediction from Eq~\ref{eq:psurv-analytical} is shown as a solid line. Other lines are guides for the eye. Parameter values: $D=16$, $K=100$, $f_R=1$ (no resistance cost), $f_S=1$ before drug addition, $f_S=0$ when antibiotic is added, $g=0.1$, $\mu=10^{-5}$, $\gamma=10^{-6}$ for the clique, $\gamma_G=3.75 \times 10^{-6}$ for the lattice, $\gamma_O=15 \times 10^{-6}$ and $\gamma_I=\alpha \gamma_O$ for the star, $\gamma_L=\gamma_R=7.5\times 10^{-6}$ for the line with $\alpha =1$, $\gamma_L=1.2\times 10^{-5}$ and $\gamma_R=\alpha \gamma_L$ for the line with $\alpha=4$. Each result is obtained from $10^4$ simulation replicates.} 
    \label{fig:latt-star-cl}
\end{figure}

\paragraph*{Different structures.} We first consider the grid or square lattice (see Fig \ref{fig:latt-star-cl}A). It is a symmetric structure, like the clique, but migrations from each deme are restricted to its four nearest neighbors, thus only allowing local migrations. In the lattice, as in the clique, in each deme the inward migrations balance the outward migrations, i.e.\ for each $i$ we have $\sum_j \gamma_{ij} = \sum_j \gamma_{ji}$, where $\gamma_{ij}$ denotes the migration rate from deme $i$ to deme $j$: this means that the lattice is a circulation~\cite{lieberman_evolutionary_2005, marrec_toward_2021,abbara_frequent_2023}. The per capita migration rate in the lattice is denoted by $\gamma_{G}$, equal in all directions. 

We further consider the star (see Fig \ref{fig:latt-star-cl}B), comprising a central deme connected to $D-1$ leaves~\cite{marrec_toward_2021}. All leaves are assumed to be equivalent. It is less symmetric than the clique or the square lattice in the sense that the central deme is different from the leaf demes. Migrations from a leaf to the center occur at per capita rate $\gamma_I$, while migrations from the center to the leaf occur at per capita rate $\gamma_O$. We define the migration asymmetry parameter as $\alpha = \gamma_I / \gamma_O$. If $\alpha=1$, the star is a circulation, while for all other values of $\alpha$ it is not, which impacts the fixation probability of a mutant in the whole population~\cite{marrec_toward_2021,abbara_frequent_2023}. 

Finally, we consider a line (see Fig \ref{fig:latt-star-cl}C). Migrations from the left to the right occur at per capita rate $\gamma_R$, from the right to the left at rate $\gamma_L$. An asymmetry parameter is defined for the line as well: $\alpha = \gamma_R/\gamma_L$. The structure is symmetric under the simultaneous transformations $\alpha \to 1/\alpha$, $\gamma_R \to \gamma_L$, and $\gamma_L \to \gamma_R$. As for the star, setting $\alpha=1$ results in a circulation, while $\alpha\ne1$ impacts the mutant fixation probability~\cite{servajean_impact_2024}.

\paragraph{Choosing migration rates.} Mutant fixation in the fastest deme is at the root of the impact of population structure on resistance evolution that we evidenced here, in particular in Fig \ref{fig:surva-prob}. Suppose that a successful R mutant appears in the fastest deme: S individuals from connected demes can migrate to the fastest deme and take over there, which takes the population back to the initial fully S state. Thus, to make different structures as comparable as possible, we set migration rates so that the rate of invasion of one deme by the connected demes is the same across structures. 

For the square lattice or grid, this leads to $\gamma_{G} =\gamma (D-1)/ 4$, where $\gamma$ is the per capita migration rate in the clique. Note that this condition also yields an equal total exchange rate in the clique and the lattice. 

For the star, we first remark that a successful mutant is $(D-1)$ times more likely to first appear in a leaf than in the center. Neglecting the case where a successful mutant first appears in the center leads to the condition $\gamma_O = (D-1)\gamma$. Note that this condition does not coincide with setting equal overall exchanges between demes in the star and in the clique. 

Finally, for the line, we note that for $D\gg 1$, successful mutants are much more likely to appear in a deme that is not at an end of the line. We thus choose migration rates such that the rate of invasion of one non-end deme by neighboring ones in the line is the same as in the clique: $\gamma_L= \gamma (D-1)/(\alpha+1)$, and $\gamma_R=\alpha\gamma_L$. Note that this condition slightly differs from setting equal overall exchanges in the line and in the clique, which would give $D$ instead of $D-1$ in the numerator of the expression of $\gamma_L$.

\paragraph{Simulation results.} Fig \ref{fig:latt-star-cl}C shows the survival probability of bacterial populations with different spatial structures with $D=16$ demes. We observe that the fully subdivided population with $\gamma=0$ has a larger survival probability than all others. This generalizes our results in Fig \ref{fig:surva-prob}. It can be attributed to the fact that S bacteria cannot re-invade the demes that fixed resistance in the absence of migrations. Moreover, we do not observe statistically significant differences between survival probabilities in the lattice and the clique with the same invasion rate. More generally, we observe that the specific graph structure of the population does not significantly affect the survival probability if the addition time $t_\textrm{add}$ of antibiotic satisfies $t_\textrm{add}\lesssim\langle t_{af F}(N^*,D)\rangle$, where $\langle t_{af F}(N^*,D)\rangle$ is the average time it takes for a successful mutant to appear in the fastest deme. The minor differences observed between the stars with different asymmetries, between them and other structures, and between lines and other structures for $t_\textrm{add}>\langle t_{af F}(N^*,D)\rangle$, can be attributed to the imperfections of our matching condition for the line and for the star, and to the fact that more than one deme may then have fixed mutants. Recall indeed that incoming migrations to the center of the star and to the end demes of the line were not matched, that our matching condition for re-invasion was constructed assuming that only one deme was mutant, and that for the star and the line, our condition did not match the condition of same total exchanges between demes.

\section{Clique population structure with mutants in the inoculum}
\label{mutino}

So far, we focused on a sensitive inoculum and on resistant individuals appearing through mutations. Let us now consider the case where mutants are already present in the inoculum. To isolate their effect, we do not include any new mutations in this case. This situation is highly relevant, especially because many experimental studies start from two bacterial strains, of which one is resistant to a given antibiotic and the other is not. In particular, a spatially structured population with mixed inoculum in each deme was recently studied~\cite{kreger_role_2023}. 

\begin{figure}[htb!]
\centering
        \includegraphics[width=0.77\linewidth]{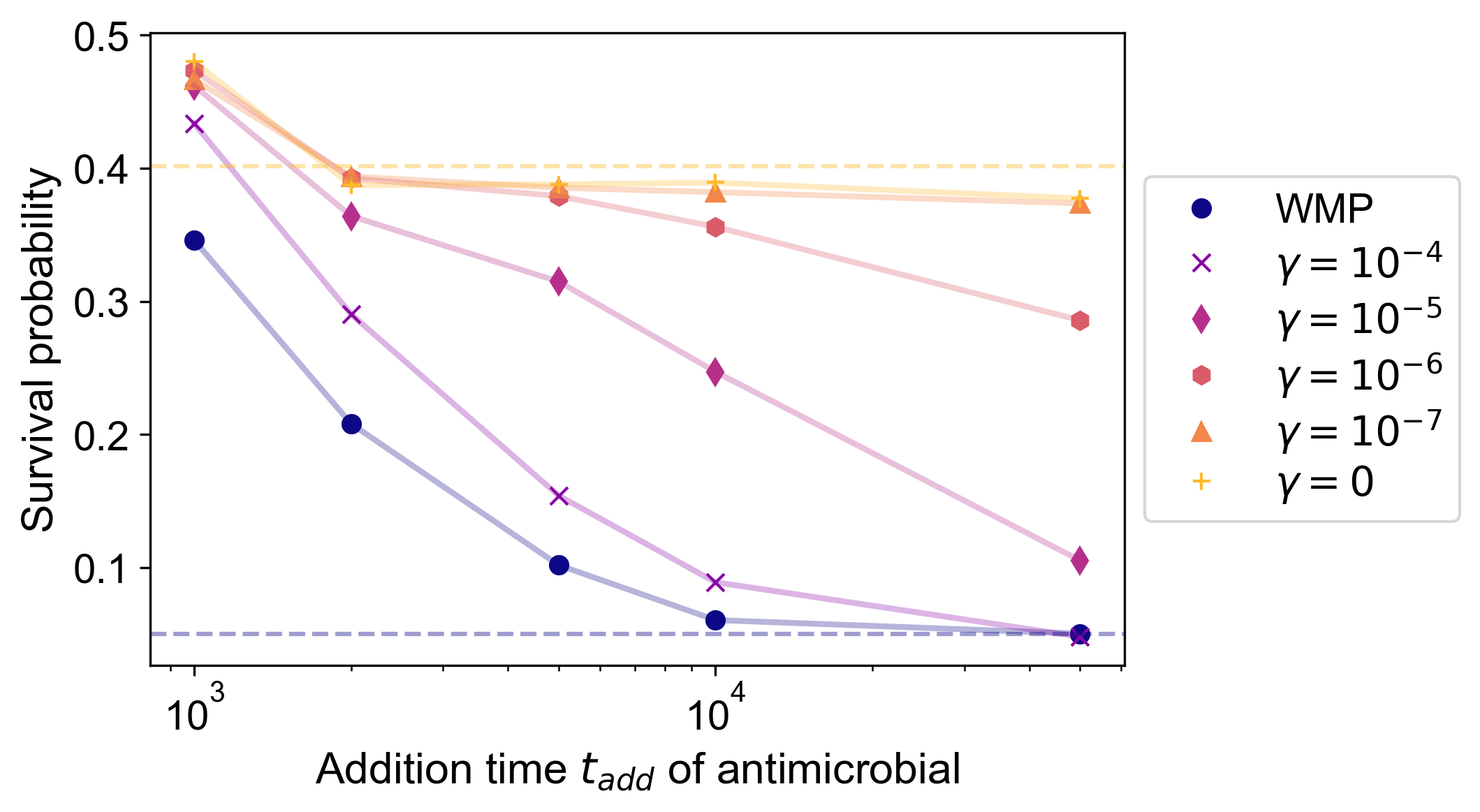}
    \caption{\textbf{Survival probability of a structured bacterial population with mutants in the inoculum upon addition of a biostatic drug.} The survival probability is plotted versus the addition time $t_\textrm{add}$ of antimicrobial. We consider a clique population structure with various migration rates. Results for a well-mixed population with same total size (``WMP") are presented as reference. Horizontal dashed lines: analytical predictions of the probabilities to still have mutants in the system in the long term (dark blue: well-mixed population; yellow: fully subdivided population with $\gamma=0$). 
    Data is obtained from $10^4$ replicate simulations in each case. Parameter values: $K=100$, $D=10$, $f_S=1$ without drug, $f_S=0$ with drug, $f_R=1$, $g = 0.1$, $\mu=0$. Initial percentage of mutants in each deme (and in the well-mixed population): 5\%.}
    \label{fig:structures-pe}
\end{figure}

In Fig \ref{fig:structures-pe}, we show the probability that a bacterial population survives biostatic antibiotic treatment versus the addition time $t_\textrm{add}$ of antibiotic for different migration rates, in the case where each deme is inoculated with 5\% mutants. Our simulation results show that the survival probability decreases when $t_\textrm{add}$ is increased in this scenario. This was expected, because new mutants do not appear, and the mutants initially present in the inoculum may go extinct. However, the spatial structure of the population still plays a crucial role. Indeed, as the migration rate $\gamma$ decreases, the survival probability increases. This is reminiscent of our result with a sensitive inoculum (see Fig \ref{fig:surva-prob}). We added the expected probabilities to still have mutants in the system as dashed lines. In a well-mixed population with initial fraction of mutants $x_\textrm{ini}$, the probability of having mutants in the long term is equal to their fixation probability, i.e.\ to $x_\textrm{ini}$ since we are considering neutral mutants. In a fully subdivided population with no migrations, this probability is equal to $1-(1-x_\textrm{ini})^D$, which is the probability that there is at least one deme where mutants fix. In a spatially structured population with nonzero migration rates, the survival probability gradually converges to the well-mixed population one because of migrations. 

\section{Concrete examples of spatially structured populations}
\label{sec:bioloex}

In hosts, different organs can feature bacterial colonization and possess spatial structures with numerous demes. Here, we provide some details on the orders of magnitudes mentioned in the Discussion.

\paragraph{Estimate of the number of intestinal crypts.} Bacteria are often found in intestinal crypts, which are glandular structures located at the base of the intestinal lining. In mice, intestinal crypts are densely packed, with approximately $10^5$ crypts in the intestine. This estimate arises from the typical distance of \SI{30}{\micro\meter} between crypts, an intestinal length of \SI{8}{\centi\meter}, and a circumference of \SI{9}{\milli\meter}~\cite{casteleyn_surface_2010}. 
Besides, an analysis of \SI{8}{\micro\meter}-thick slices of mouse colon crypts revealed the presence of 15 to 35 bacteria in each slice~\cite{pedron_acrypt_2012}. Considering a crypt depth of approximately \SI{100}{\micro\meter}~\cite{dekaney_expansion_2007}, this yields a range of 150 to 450 bacteria per crypt. 
Note that a given infection may not affect all crypts in a host. 

\paragraph{Estimate of the number of skin pores.} Bacteria also colonize skin pores in humans~\cite{conwill_anatomy_2022}. In the skin of the human face and nose, there are approximately 20 to 30 pores per \SI{0.8}{\cm\squared}~\cite{campos_useof_2019}. With the average area of the face being around \SI{600}{\cm\squared}, this corresponds to roughly \SI{2e4}{} pores.

\addtocontents{toc}{\protect\setcounter{tocdepth}{0}}
\printbibliography[heading=bibintoc]

\end{document}